\documentclass[journal]{IEEEtran}
\usepackage{amsmath,amsfonts}
\usepackage{algorithmic}
\usepackage{algorithm}
\usepackage{array}
\usepackage[caption=false,font=normalsize,labelfont=sf,textfont=sf]{subfig}
\usepackage{textcomp}
\usepackage{stfloats}
\usepackage{url}
\usepackage{verbatim}
\usepackage{graphicx}
\usepackage{cite}
\usepackage{amssymb}
\usepackage{bbding}
\usepackage{booktabs}
\usepackage{color}
\usepackage{diagbox}
\usepackage{makecell}
\usepackage{multirow}
\usepackage{pifont}
\usepackage[colorlinks]{hyperref}
\newcommand{\yesyes}{\ding{51}}

\usepackage{fontawesome}

\begin{document}

\title{
Vision Calorimeter for High-Energy Particle Detection
}

\author{
    Hongtian Yu~\textsuperscript{\faEnvelopeO},
    Yangu Li,
    Yunfan Liu,
    Yunxuan Song,
    Xiao-Rui Lyu,
    Qixiang Ye$^{\dagger}$~\textsuperscript{\faEnvelopeO}
\thanks{$^\dagger$Corresponding author.}
\thanks{\textsuperscript{\faEnvelopeO}E-mail: yuhongtian17@mails.ucas.ac.cn, qxye@ucas.ac.cn.}
}

\markboth{Manuscript for Review}%
{Shell \MakeLowercase{\textit{et al.}}: A Sample Article Using IEEEtran.cls for IEEE Journals}


\maketitle

\begin{abstract}

In high-energy physics, estimating anti-neutron parameters (position and momentum) using the electromagnetic calorimeter (EMC) is crucial but challenging.
To conquer this challenge, we propose Vision Calorimeter (ViC), a framework that migrates visual object detectors to analyze particle images.
The motivation lies in introducing a physics-inspired heat-conduction operator (HCO) into the detector's backbone and head to handle the discrete and sparse patterns of these images.
Implemented via the Discrete Cosine Transform, HCO extracts frequency-domain features, bridging the distribution gap between natural and particle images.
Experiments demonstrate that ViC significantly outperforms conventional methods, reducing the incident position prediction error by 46.16\% (from 17.31° to 9.32°) and providing the first baseline result with an incident momentum regression error of 21.48\%.
This study underscores ViC's great potential as a reliable particle detector for high-energy physics.
Code is available at \href{https://github.com/yuhongtian17/ViC}{https://github.com/yuhongtian17/ViC}.

\end{abstract}


\section{Introduction}

Particle physics explores the most fundamental building blocks of the natural world and the forces that govern their interactions.
A key experimental apparatus in this field is the \textit{collider}, where two particles are accelerated to pseudo-light speeds and collide head-on.
Such collisions generate a diverse array of particles, which decay over time and are subsequently detected by sensors, $e.g.$, calorimeters~\cite{Fabjan:2003aq}, positioned around the collision point.
Estimating the properties of the decayed final-state particles enables researchers to measure the initial-state particles and their decay process accurately, further compare them with theoretical values and make analyses.
Among the final-state particles, \textit{anti-neutrons} ($\bar{n}$) stand out as an important category.

However, due to the absence of detecting materials for nuclear-nuclear interactions within certain energy ranges and the electrical neutrality of $\bar{n}$, researchers can only rely on the electromagnetic calorimeter (EMC) to capture partial energy information.
Such information represents discreteness and sparsity, making it challenging to utilize for \textbf{parameter estimation} (also expressed as \textit{reconstruction} in physics), which includes \textbf{incident position prediction} and \textbf{incident momentum regression}.
Conventional methods, such as those based on analytical clustering algorithms~\cite{He:2011zzd}, struggle to reliably distinguish $\bar{n}$ from other particles and noise, not to mention accurately determine their position and momentum.
Recognizing the capability of large-scale deep models to process vast amounts of data and extract underlying patterns, we propose employing these models to estimate $\bar{n}$ status.
Since this task fundamentally involves predicting both the location (position) and the label (momentum) of $\bar{n}$, it can naturally be formulated as an \textbf{object detection} problem in computer vision.
Consequently, the primary challenge lies in migrating visual object detectors to effectively process particle images derived from high-energy physics experiments.

\begin{figure*}[!t]
    \centering
    \includegraphics[width=0.9\linewidth]{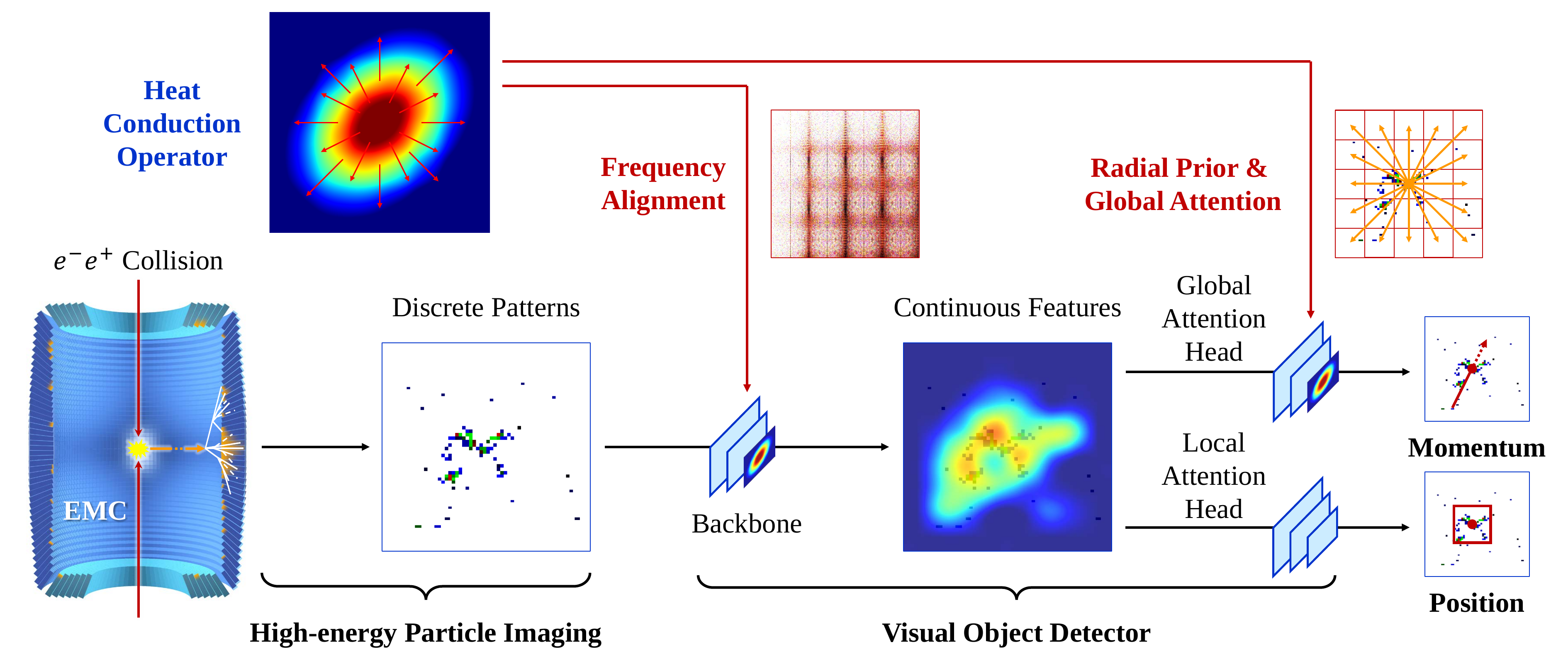}
    \caption{
    Overview of the Vision Calorimeter (ViC) pipeline. ViC integrates a novel heat-conduction operator into both the backbone and head of a visual object detector. 
    Inspired by the principles of physical heat-conduction, this operator combines a radial prior with global attention, effectively capturing the unique characteristics of particle patterns. Implemented via discrete cosine transform (DCT), it ensures seamless alignment with pre-trained visual representation, enabling efficient transfer learning from natural images to particle images. (Best viewed in color)
    }
    \label{fig:pipeline}
\end{figure*}

In this study, we propose Vision Calorimeter (ViC), a framework based on visual object detectors specifically adapted for the parameter estimation of $\bar{n}$, Fig.~\ref{fig:pipeline}.
To ensure compatibility of data representation, we establish a correspondence between spatial arrangement of the EMC cells and coordinates of the image pixels, mapping the EMC surface onto the image plane.
By further quantifying the EMC recordings and encoding them as RGB values, we construct a visual representation of the final-state particles, which we refer to as \textbf{high-energy particle image}.
A preliminary examination of such particle images (Fig.~\ref{fig:pipeline}) reveals substantial differences from natural images: (1) discrete and sparse patterns and (2) scattered regions of foreground activation.

For these characteristics, we carry out a series of adaptations and optimizations to facilitate the migration of visual object detectors.
\textbf{First}, we employ the heat-conduction operator (HCO)~\cite{vheat2025}, a physics-inspired visual representation module based on 2-D Discrete Cosine Transform (DCT), as core operator of the backbone network, Fig.~\ref{fig:pipeline}.
We further improve the original HCO by making heat conductivity coefficient dependent on input samples.
This module extracts frequency-domain features, mitigating the challenges caused by discrete and sparse patterns, and facilitating aligning with the pre-trained visual representation.
\textbf{Second}, to align with the visual object detection pipeline, we propose an annotation strategy to generate pseudo bounding boxes.
\textbf{Third}, we take the radial prior and global attention advantage from HCO and improve structure of the detector head to better address this issue.
This is based on the observation that there exists conflict between local attention mechanism and scattered nature of the deposited energy for momentum regression.
We further introduce two metrics, \textit{i.e.}, the mean angular bias (mAB) for position prediction and the mean relative error (mRE) for momentum regression, guided by physics practices to evaluate the accuracy of particle detection.
The final detection framework, noted as ViC, achieves 9.32° mAB and 21.48\% mRE, reducing the incident position prediction error by 46.16\% compared with conventional clustering algorithms (17.31° mAB).

The contributions of this study are summarized as:
\begin{itemize}
    \item We introduce Vision Calorimeter (ViC), the first end-to-end deep learning baseline for anti-neutron detection based on EMC data, through migrating visual object detectors to high-energy particle images.

    \item 
    We format the high-energy particle image and introduce the heat-conduction operator (HCO) to the visual object detector, specifically tailored to the discrete and scattered patterns of the particle incidence.

    \item Experimental results demonstrate that ViC not only significantly outperforms the conventional method in incident position prediction but also enables measurement of incident momentum, for the first time.
    This highlights its potential in modeling the extensive data generated by high-energy colliders.
\end{itemize}

\section{Related Work}

\subsection{Machine Learning in Particle Physics}

The application of machine learning methods in particle physics has a long history~\cite{Bowser-Chao:1992giy}.
Commonly used multivariate analysis techniques~\cite{Voss:2007jxm}, such as decision trees, support vector machines (SVMs), and artificial neural networks (ANNs), have been widely adopted.
For instance, the discovery of the Higgs boson~\cite{ATLAS:2012yve, CMS:2012qbp} was aided by decision trees, which were used to classify signal and background collision events.
While these methods are lightweight and easy to implement, their performance is constrained by the shallow architectures and hand-crafted feature representations.

Recently, the powerful representation capabilities of deep learning models, such as convolutional neural networks (CNNs)~\cite{lenet, resnet}, Transformers~\cite{attention, vit}, and graph neural networks (GNNs)~\cite{scarselli2008graph, qu2020jet, qu2022particle}, have inspired novel approaches to revolutionize particle physics experiments~\cite{wang2023scientific}.
These advances have been applied in areas such as simulating detector responses~\cite{Hashemi:2023rgo}, reconstructing generated particles~\cite{Duarte:2020ngm}, and analyzing physics objects~\cite{Mondal:2024nsa}.
Despite the progress, deep learning methods for particle \textit{reconstruction} remain underdeveloped.
On the one hand, commonly used methods in practice~\cite{Qasim:2019otl}, which rely on step-wise strategies, $e.g.$, clustering followed by classification, are not only complex but also sub-optimal.
On the other hand, subsequent studies investigate some end-to-end formulations for this problem~\cite{komiske2019energy, Kieseler:2020wcq} where position prediction, momentum regression and particle identification in a unified framework, but cannot meet practical requirements.
Especially, there does not exist a(n) (anti-)neutron \textit{reconstruction} method which uses solely an EMC.
Physicists usually adopted a simple analytical clustering algorithm~\cite{He:2011zzd} to detect photons according to the deposited energies on the EMC, but for neutrons, the uncertainty of both position and momentum with this algorithm is quite large.

\subsection{Visual Representation}

Visual representation is of vital significance in image processing.
It has gone through a long period of hand-crafted feature design, during which the wavelet transform~\cite{Wavelet1989}, SIFT~\cite{sift}, HOG~\cite{hog} are particularly outstanding, and has been strongly impacted by deep learning models in recent years.
CNNs have been an early proposed and widely used deep representation model~\cite{lenet} and demonstrate powerful performance after years of polishing~\cite{resnet, convnext}.
Transformer was initially proposed as a model for natural language processing~\cite{attention}, and has since been adapted for use in visual tasks, where it continues to demonstrate strong representational capabilities~\cite{vit, swin}.
More recently, researchers have explored a number of alternative representation models that are not based on CNNs or Transformers~\cite{vim, vmamba, vheat2025}.
However, the involved objects are mostly confined to natural scene images, which are easy to be collected from the internet.
By contrast, discrete particle images have very different characteristics -- such as sparsity and radial patterns -- to natural scene images.
To find out plausible feature representation for such patterns is the key to build the Vision Calorimeter.

\subsection{Visual Object Detection}

\begin{table*}[!t]
\centering
\setlength{\tabcolsep}{0.20cm}
\caption{Correspondence relationship between EMC cells and image pixels.}
\begin{tabular}{c|ccc|ccc|c|ccccccc|c|ccc|ccc}
\toprule
note    & \multicolumn{3}{c|}{gap} & \multicolumn{3}{c|}{end-cap} 
        & gap 
        & \multicolumn{7}{c|}{barrel} 
        & gap 
        & \multicolumn{3}{c|}{end-cap} & \multicolumn{3}{c}{gap} \\
\midrule
layers  & 2 & 2 & 3 & 2 & 2 & 2 
        & 1 
        & 5 & 4 & 5 & 16 & 5 & 4 & 5 
        & 1 
        & 2 & 2 & 2 & 3 & 2 & 2 \\
cells   & - & - & - & 64 & 80 & 96 
        & - 
        & 120 & 120 & 120 & 120 & 120 & 120 & 120 
        & - 
        & 96 & 80 & 64 & - & - & - \\
\midrule
w (pixels) & 30 & 24 & 20 & 15 & 12 & 10 
        & 10 
        & 8 & 8 & 8 & 8 & 8 & 8 & 8 
        & 10 
        & 10 & 12 & 15 & 20 & 24 & 30 \\
h (pixels) & 8 & 8 & 7 & 6 & 6 & 5 
        & 5 
        & 5 & 6 & 7 & 8 & 7 & 6 & 5 
        & 5 
        & 5 & 6 & 6 & 7 & 8 & 8 \\
\bottomrule
\end{tabular}
\label{tab:cells}
\end{table*}

\begin{figure}[!t]
    \centering
    \includegraphics[width=1.0\linewidth]{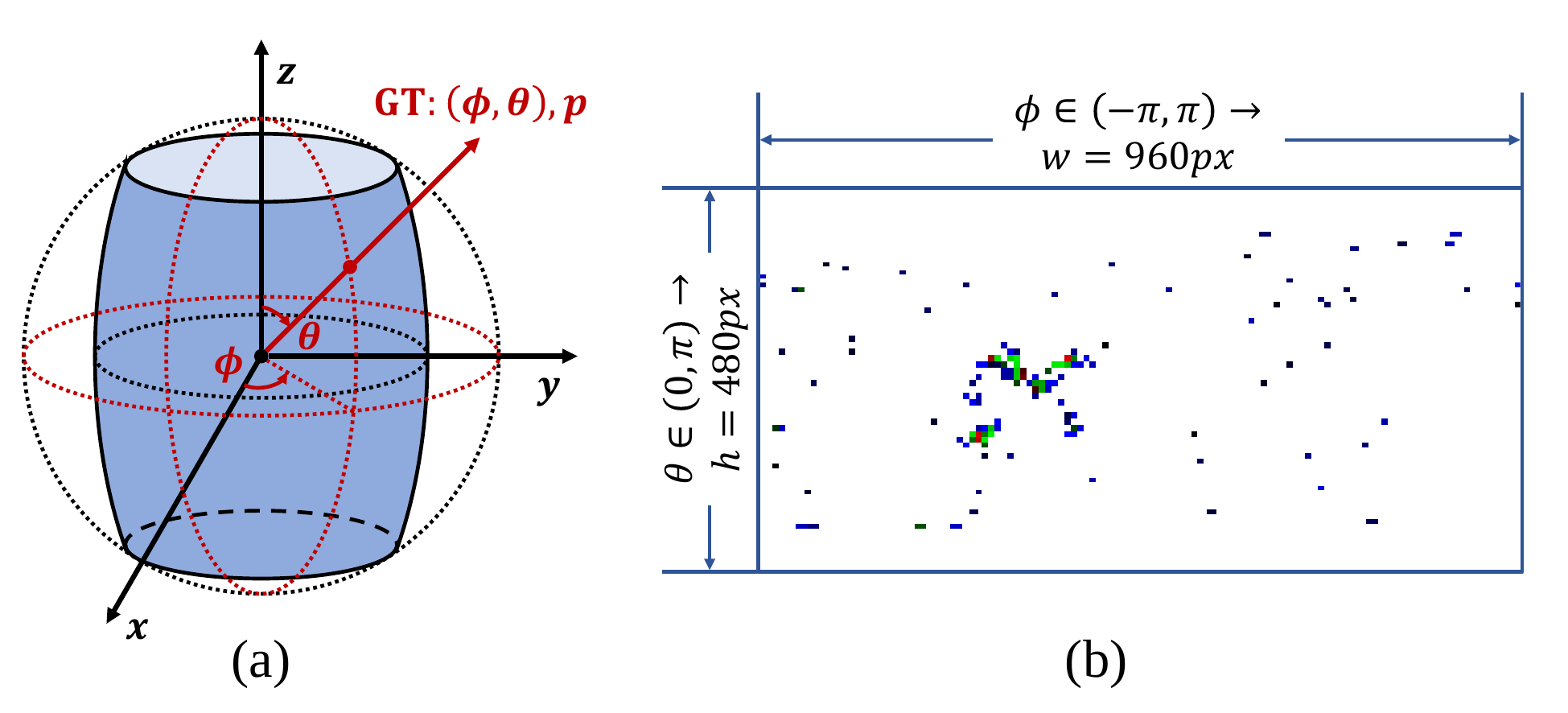}
    \caption{
    Formatting a high-energy particle image.
    (a) EMC cell arrays are arranged in a spherical coordinate system to capture the spatial structure of particle interactions.
    (b) The energy depositions are projected to a 2-D image domain, representing the readout as pixel intensities for subsequent analysis.
    }
    \label{fig:particle_image}
\end{figure}

\begin{figure}[!t]
    \centering
    \includegraphics[width=1.0\linewidth]{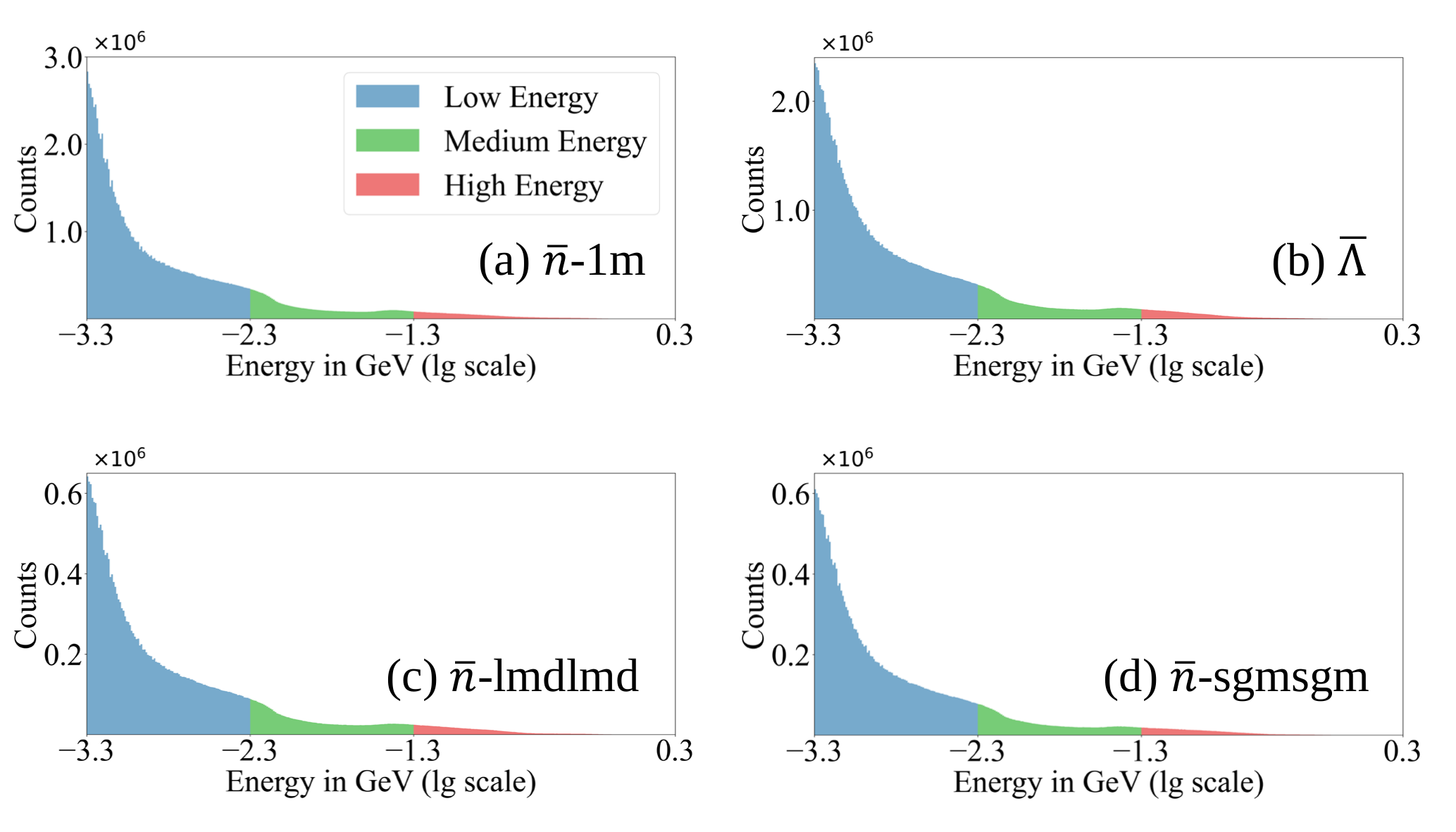}
    \caption{
    Histogram of deposited energy recorded by EMC.
    }
    \label{fig:eng2rgb}
\end{figure}

This is a computer vision approach to determine the position $(x, y)$ and size $(w, h)$ of a given object within an image.
Taking advantages of the deep learning models as backbones, modern detectors were endowed powerful capability to precisely identify and localize objects of interests from complex and noisy backgrounds~\cite{voc, mscoco}.
Compared with image classifiers, object detectors enjoy two additional advantages:
(i) the backbone's representation capability is decoupled to perform classification and localization tasks at the same time~\cite{fasterrcnn, retinanet}, and
(ii) the classifier is optimized to handle the background-foreground imbalance~\cite{freeanchor, paa}.
The first advantage endows the capability to perform anti-neutron parameter estimation which requires to perform incident position prediction and particle classification at the same time.
The second one facilitates solving the problem of unbalanced signal and background events.

Researchers have customized styles of object detectors to meet different needs.
For example, the YOLO series~\cite{yolov1, yolov12} are developed for real-time object detection;
single-stage object detectors~\cite{ssd, retinanet} are developed with simple structure and good scalability;
two-stage object detectors~\cite{fasterrcnn, maskrcnn} are suitable for dense detection with high precision;
end-to-end object detectors~\cite{detr, dino} take out NMS post-processing.
These methods exhibit strong generalization capacity for object detection in natural scene images, but remains to be elaborated for high-energy particle images.

\section{Preliminary: Formatting High-energy Particle Images}

To estimate the parameters of $\bar{n}$ produced in particle collision events using visual object detectors, it is required to collect the deposited energy readouts from the EMC cells and format these signals as 2-D images, Fig.~\ref{fig:particle_image}.

\subsection{Collision Event Data Collection}

To build the Vision Calorimeter, we first collect a dataset with electron-positron collision events upon the BESIII experiment~\cite{BESIII:2022rgl}.
The events in the dataset span a momentum range of $\bar{n}$ from 0 to 1.2 GeV/$c$, which sufficiently covers the needs of most physical analysis at BESIII.
Experimental analysis confirmed that over 99\% of the events contain a single $\bar{n}$~\cite{BES:2004gwe}.

In the dataset, raw data of a single event includes an azimuthal angle ($\phi$), a polar angle ($\theta$), an energy readout (in GeV) for each activated EMC cell, which is triggered either by the incident $\bar{n}$ or by background noise.
For ground-truth labeling, the azimuthal angle ($\phi\in(-\pi, \pi)$), polar angle ($\theta\in(0, \pi)$), and momentum magnitude ($p\in(0, 1.2)$ GeV/$c$) of the incident $\bar{n}$ is derived using the energy-momentum conservation law, Fig.~\ref{fig:particle_image}(a).
The labeling procedure introduces inherent uncertainties of 0.61° mAB and 0.74\% mRE~\footnote{Please refer to the \textit{Evaluation Metric} subsection for definitions of these evaluation metrics.}.
In specific physics scenarios, $e.g.$, $J/\psi{\rightarrow}p{\pi}^{-}\bar{n}$, the ground-truths can be mathematically calculated by recoiling against other well-reconstructed particles.
In more general cases, $e.g.$, $\bar{\Lambda}_{c}^{-}{\rightarrow}\bar{n}e^{-}\bar{\nu}_e$~\cite{BESIII:2024mgg}, the ground-truths cannot be pre-determined in this way due to unmeasurable particles like neutrino $\bar{\nu}_e$.
The significance of this study is to capture visual patterns of $\bar{n}$ images from the labeled dataset and extend these insights to general cases, thereby advancing further physics research.

\subsection{Visual Representation of Collision Event}

\subsubsection{Cell Arrays to Image Pixels}

As shown in Fig.~\ref{fig:pipeline}, EMC cells are arranged on a cylindrical surface comprising a barrel section and two end-caps. Geometric gaps exist between the barrel and the end-caps, as well as between the end-cap and the pole.
The barrel section consists of 44 circular layers, each with 120 cells distributed uniformly along the circumference, where each cell spans an angular width of 3°.
The end-caps consist of 6 layers and the cell number per layer decreases from the outermost to the innermost: 96, 96, 80, 80, 64, 64.

To represent this arrangement as a 2-D unwrapped image, the image width ($w_{img}$) is set to 960 pixels (the least common multiple of the cell counts) to that each cell map be an integer number of pixels.
The image height ($h_{img}$) is set to 480 pixels, maintaining the 2:1 ratio between the azimuthal angle ($\phi \in (-\pi, \pi)$) and the polar angle ($\theta \in (0, \pi)$).
The correspondence between spherical coordinates $(\phi, \theta)$ and the 2-D image coordinates $(x, y)$ is calculated as
\begin{equation}
    \begin{aligned}
    \frac{x}{w_{img}} & =\frac{\phi+\pi}{2\pi} & \in(0, 1), \\
    \frac{y}{h_{img}} & =\frac{\theta}{\pi} & \in(0, 1). \\
    \end{aligned}
    \label{eq:xy2phithe}
\end{equation}

To make cell arrays within the EMC surface strictly correspond to 2-D image pixels, Table~\ref{tab:cells}, we take two measures:
(i) For each cell in barrel and end-cap regions, we calculate the pixel number in the image through a proportional relationship;
(ii) For the geometric gaps, we perform an interpolation.

\subsubsection{Deposited Energy to Pixel Intensity}

\begin{figure}[t]
    \centering
    \includegraphics[width=0.95\linewidth]{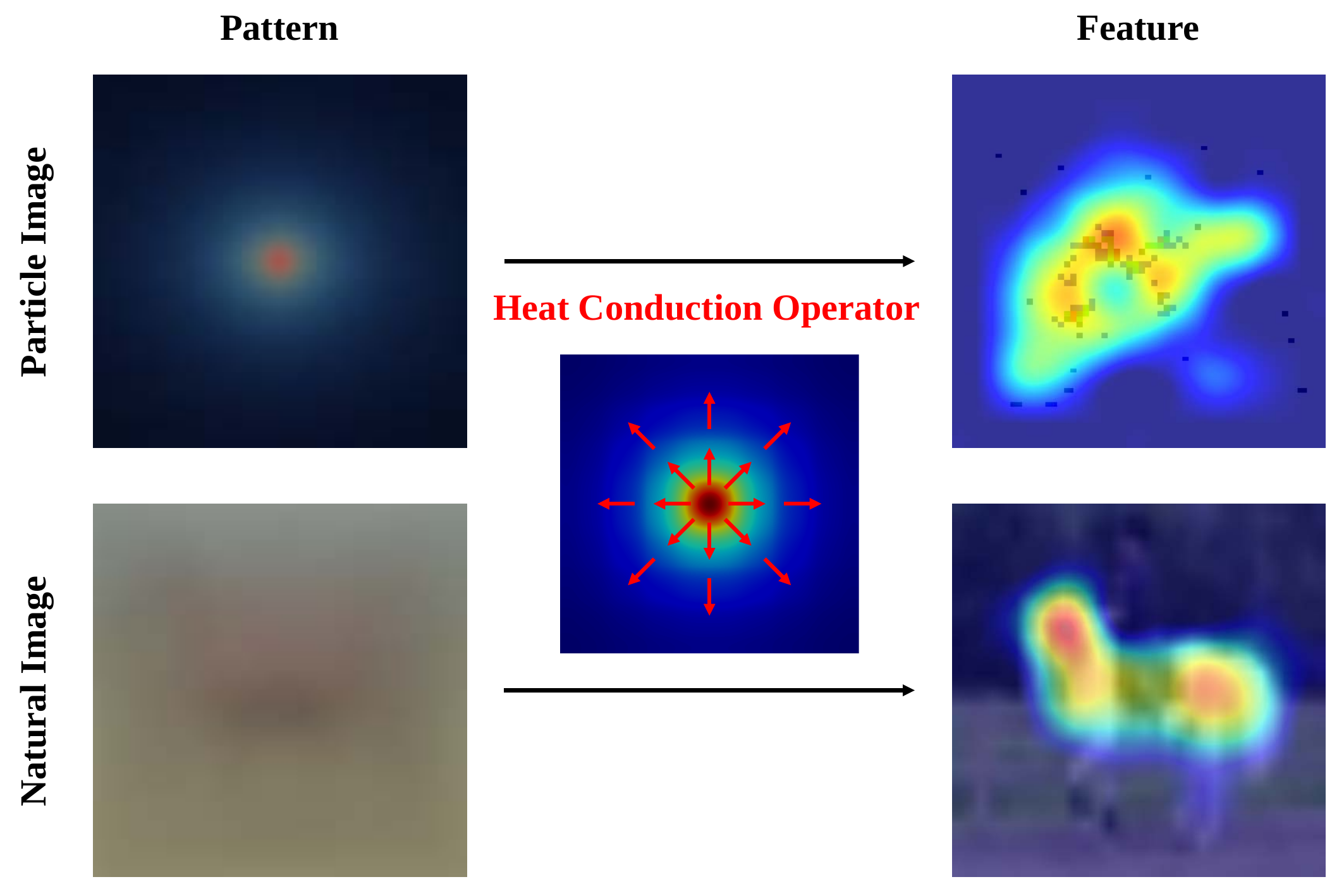}
    \caption{
    The average particle image manifests the energy deposition pattern of $\bar{n}$. After extraction by the heat-conduction operator, the features of particle images become akin to those of natural images. (The horse image shown here is sourced from the CIFAR-10 dataset~\cite{cifar}.)
    }
    \label{fig:heat}
\end{figure}

The EMC cells record the energy deposited by an incident particle. When mapping EMC cells to 2-D image pixels, Fig.~\ref{fig:particle_image}(b), the EMC energy readouts are naturally mapped to the intensity values of corresponding pixels.
Unfortunately, the EMC energy distribution is highly imbalanced, $i.e.$, most values lie in the low-energy range, Fig.~\ref{fig:eng2rgb}.
To address this, we propose to divide the range of readouts ($E \in [5 \times 10^{-4}, 2] \approx [10^{-3.3}, 10^{0.3}]$ GeV) to three intervals:
\textit{low energy} (lower than $10^{-2.3}$ GeV),
\textit{medium energy} (between $10^{-2.3}$ and $10^{-1.3}$ GeV), and
\textit{high energy} (higher than $10^{-1.3}$ GeV),
Energy values within these intervals are respectively map to gray values in the blue (B), green (G), and red (R) color channels within the RGB color space (named energy-to-color mapping).
To reduce the scope of energy values, we first apply a logarithmic transformation (base 10) as pre-processing.
We then apply the image histogram equalization~\cite{pizer1987HistEqual} to alleviate the impact of extreme values.
The above pre-processing procedure is formulated as
\begin{flalign}
    \frac{B}{255} 
                  &\ = (\lg{E}+3.3)^{0.5} &
                  &\ E\in(-\infty, 10^{-2.3}), & \nonumber \\
    \frac{G}{255} 
                  &\ = (\lg{E}+2.3)^{0.6} &
                  &\ E\in[10^{-2.3}, 10^{-1.3}), & \nonumber \\
    \frac{R}{255} 
                  &\ = \frac{\arctan{((\lg{E}+1.3)\cdot2.5)}}{\arctan{3}} &
                  &\ E\in[10^{-1.3}, +\infty). & \nonumber \\
    \label{eq:eng2rgb}
\end{flalign}

\section{Approach: Parameter Estimation with Visual Object Detectors}

Given particle images derived from EMC collision events, ViC predicts the incident position and the incident momentum of $\bar{n}$ by migrating a visual object detector, Fig.~\ref{fig:pipeline}.

\subsection{Representing Discrete Particle Patterns}

Unlike natural images which feature strong textural and semantic continuity, particle images are characterized by greater randomness and fragmentation.
The foreground exhibits discrete patterns, while the majority of the image consists of blank and uninformative areas, highlighting its sparsity.
These unique characteristics pose challenges for existing visual representation models.

By observation, Fig.~\ref{fig:heat}, we find that the energy characteristic of incident point centrality and isotropic radial diffusion revealed in the average particle image strongly resembles the physical heat-conduction process.
Inspired by this similarity, we adopt the heat-conduction operator (HCO)~\cite{vheat2025}, a physics-driven representation module that models the feature extraction process by drawing an analogy to physical heat-conduction.
Specifically, HCO maps complex features to high-temperature regions where heat accumulates, and sparse features to low-temperature regions where heat dissipates more readily.
Using the discrete solution of the heat-conduction equation, HCO performs forward computation of features in a manner analogous to the physical heat diffusion.

We utilize HCO implementation via the 2-D Discrete Cosine Transform (DCT) and its inverse (IDCT).
In the spatial domain, particle images are typically represented as discrete pulses or rectangular window signals.
Through the Fourier Transform, these signals are transformed into continuous representations in the frequency domain, making the features of particle images similar to those of natural images, Fig.~\ref{fig:heat}.
The spatial-frequency transform bridges the domain gap between two types of images, thereby simplifying the modeling of discrete patterns and improving compatibility with detectors pre-trained on natural images.
Specifically, the HCO in each layer is calculated as
\begin{equation}
    U^t = \mathbf{\mathrm{IDCT}}_\mathbf{\mathrm{2D}}\left( \mathbf{\mathrm{DCT}}_\mathbf{\mathrm{2D}}(U^0)e^{-k(\omega_x^2 + \omega_y^2)t} \right),
\end{equation}
where $(x, y)$ denotes the spatial domain coordinates while $(\omega_x, \omega_y)$ denotes the frequency domain coordinates. $U^t$ denotes a feature map after heat-conduction with duration $t$. $k$ denotes the heat conductivity~\footnote{Please refer to the \textit{Appendix} for a detailed derivation of the heat-conduction process using DCT/IDCT transforms.}.

\begin{figure}[!t]
    \centering
    \includegraphics[width=0.95\linewidth]{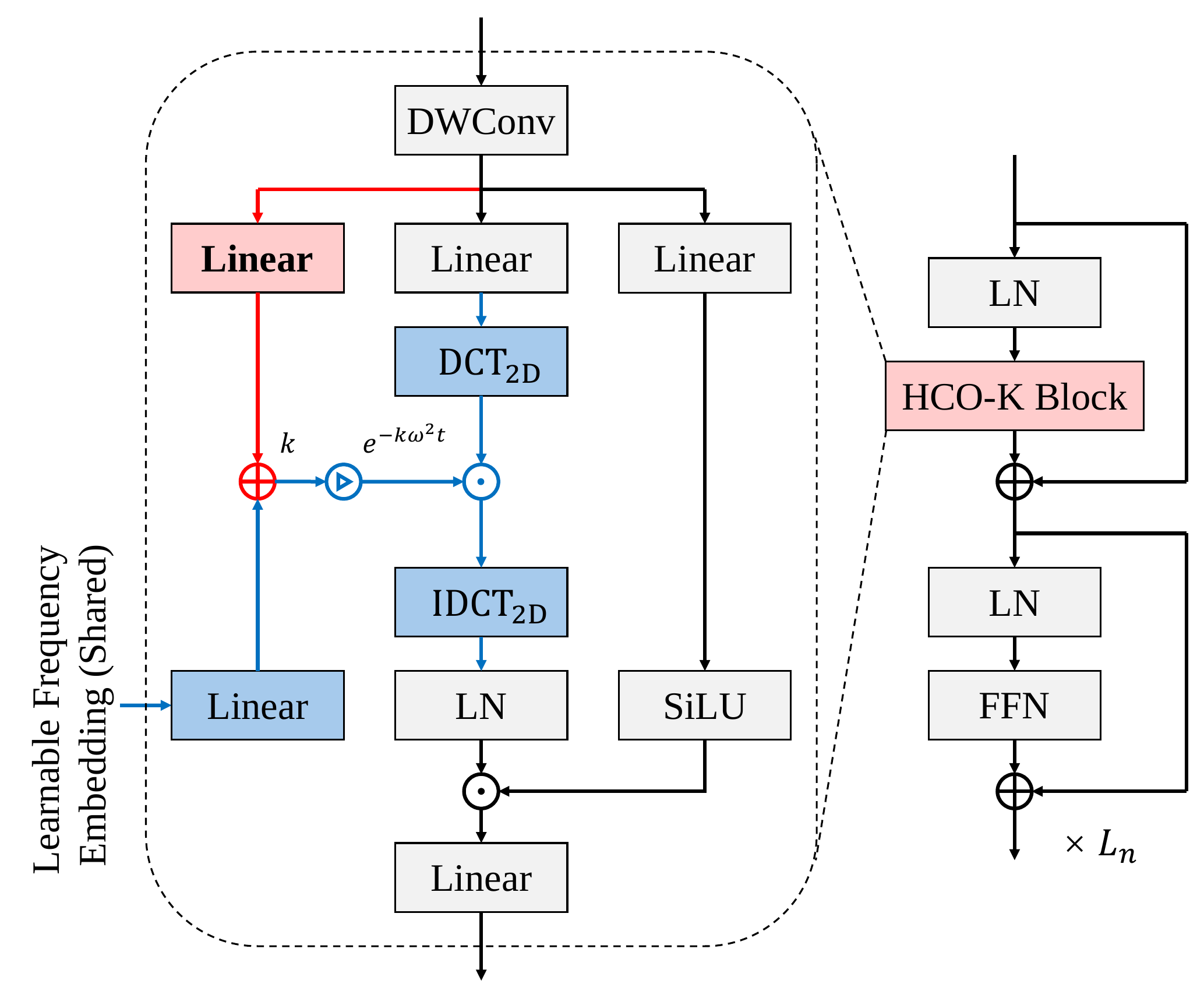}
    \caption{
    Heat-conduction operator (HCO), which integrates blocks of layer normalization (LN), feed-forward network (FFN), and depth-wise convolution (DWConv). Blocks in blue represent features of the original HCO design, while elements in red highlight the modifications introduced in HCO-K.
    }
    \label{fig:hco}
\end{figure}

To better adapt the visual representation model to particle images, we introduce technical improvements to the calculation of heat conductivity $k$ in HCO.
In its original design, $k$ is computed using a learnable frequency embedding shared across samples and layers, effectively capturing common conduction characteristics.
However, this design overlooks the inherent differences between samples and layers, which is particularly pronounced in particle image datasets compared to natural image datasets.
To address this, we preserve the original $k$ while making it dependent on the feature map of each layer through feature fusion.
This modification retains the shared characteristics while dynamically capturing the distinct conduction attributes of different samples and layers.
We name the improved module HCO-K.

Drawing inspiration from the way HCO is used to build vHeat~\cite{vheat2025}, HCO-K adopts a similar approach to construct HeatK Layers (as illustrated in Fig.~\ref{fig:hco}).
These HeatK Layers are stacked following the hierarchical design principles of the Swin Transformer~\cite{swin}, forming the final backbone network, which is named vHeatK.
The network incorporates fundamental components such as linear blocks, layer normalization, and feed-forward networks, which are directly derived from the design of Transformer~\cite{attention}.
In addition, a depth-wise convolution is employed to smooth the extracted features, while a SiLU branch is introduced to enhance the model's nonlinear representation capabilities and regulate the flow of feature information.

\subsection{Implementing Particle Detection}

\subsubsection{A Straight-forward Solution}

An intuitive approach of $\bar{n}$ detection is to directly predict the labels from the particle image.
To achieve this, we design a \underline{s}traight-\underline{f}orward \underline{net}work (SFNet) to regress the incident position $(\phi_{pred}, \theta_{pred})$ and the incident momentum $p_{pred}$ by predicting several scale factors from the unwrapped image.
Specifically, SFNet employs a deep neural network as the backbone, a simple FC-Sigmoid layer as the head network, and the binary cross-entropy as the loss function to predict the fractional values $(\lambda_1, \lambda_2, \lambda_3)$, as
\begin{equation}
    \begin{aligned}
    \lambda_1 & =\frac{x_{pred}}{w_{img}}=\frac{\phi_{pred}+\pi}{2\pi} & \in(0, 1), \\
    \lambda_2 & =\frac{y_{pred}}{h_{img}}=\frac{\theta_{pred}}{\pi} & \in(0, 1), \\
    \lambda_3 & =\frac{p_{pred}}{p_{max}} & \in(0, 1), \\
    \end{aligned}
\end{equation}
where $p_{max}$ is set to 1.2 GeV/$c$.
For mAB and mRE metrics, the state-of-the-art deep learning model, $e.g.$, our proposed vHeatK, achieves only a modest improvement of 4.44° over the clustering-based algorithm~\cite{He:2011zzd} (referred to as ClustAlgo).
This result highlights the necessity of developing a more effective detection model.

\subsubsection{Incident Position Prediction}

\begin{table}[t]
\centering
\setlength{\tabcolsep}{0.075cm}
\caption{Performance comparison of ClustAlgo and SFNet using different backbone models.}
\begin{tabular}[t]{@{}c|c|ccccc@{}}
\toprule
& \multirow{2}*{ClustAlgo} & \multicolumn{5}{c}{SFNet} \\
& & ResNet-50 & ConvNX-T & Swin-T & vHeat-T & vHeatK-T \\
\midrule
mAB (°)  $\downarrow$ & 17.31 & 24.71 & 14.70 & 15.50 & 13.07 & \textbf{12.87} \\
mRE (\%) $\downarrow$ & -     & 31.58 & 23.90 & 23.35 & 22.11 & \textbf{22.04} \\
\bottomrule
\end{tabular}
\label{tab:abla_sfnet_backbones}
\end{table}

\begin{figure}[t]
    \centering
    \includegraphics[width=1.0\linewidth]{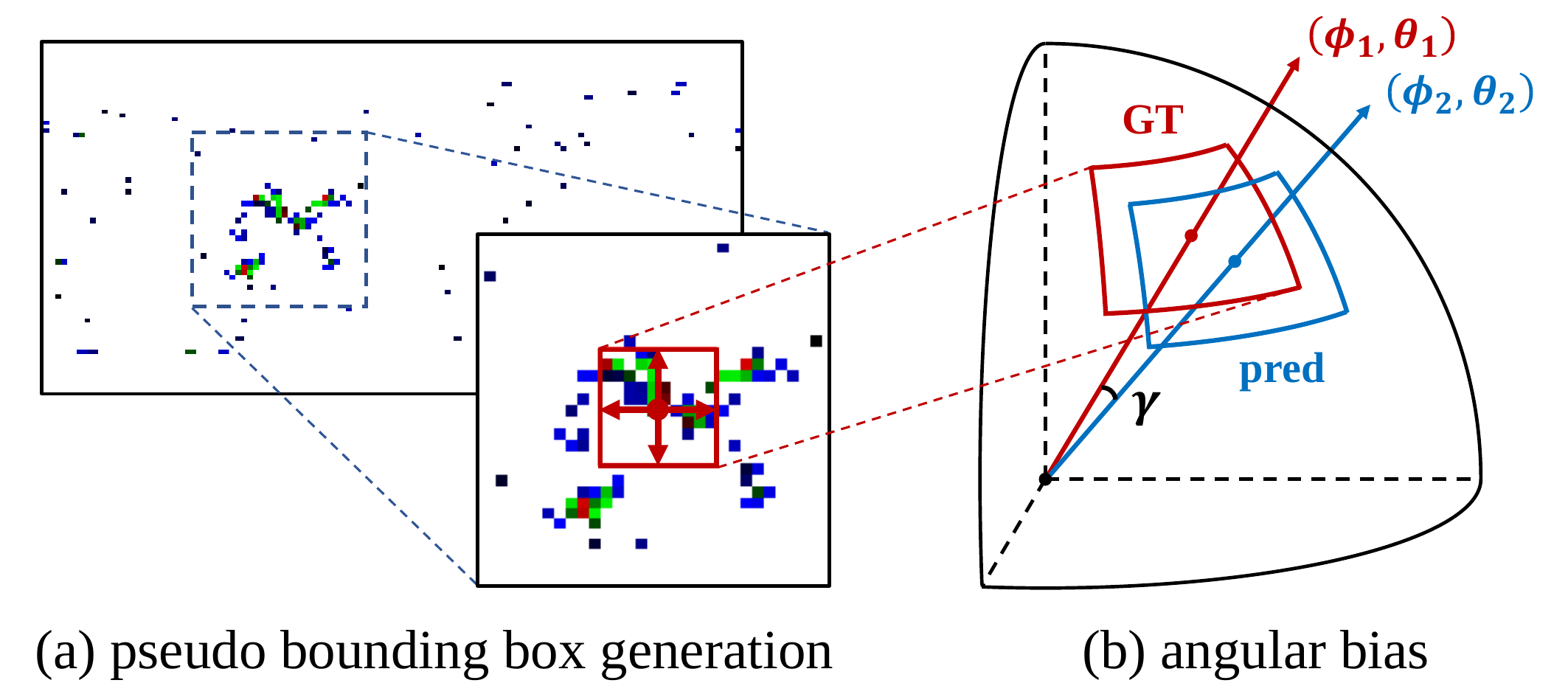}
    \caption{Illustration of incident position prediction.}
    \label{fig:pseudo_bbox}
\end{figure}

\begin{figure}[t]
    \centering
    \includegraphics[width=1.0\linewidth]{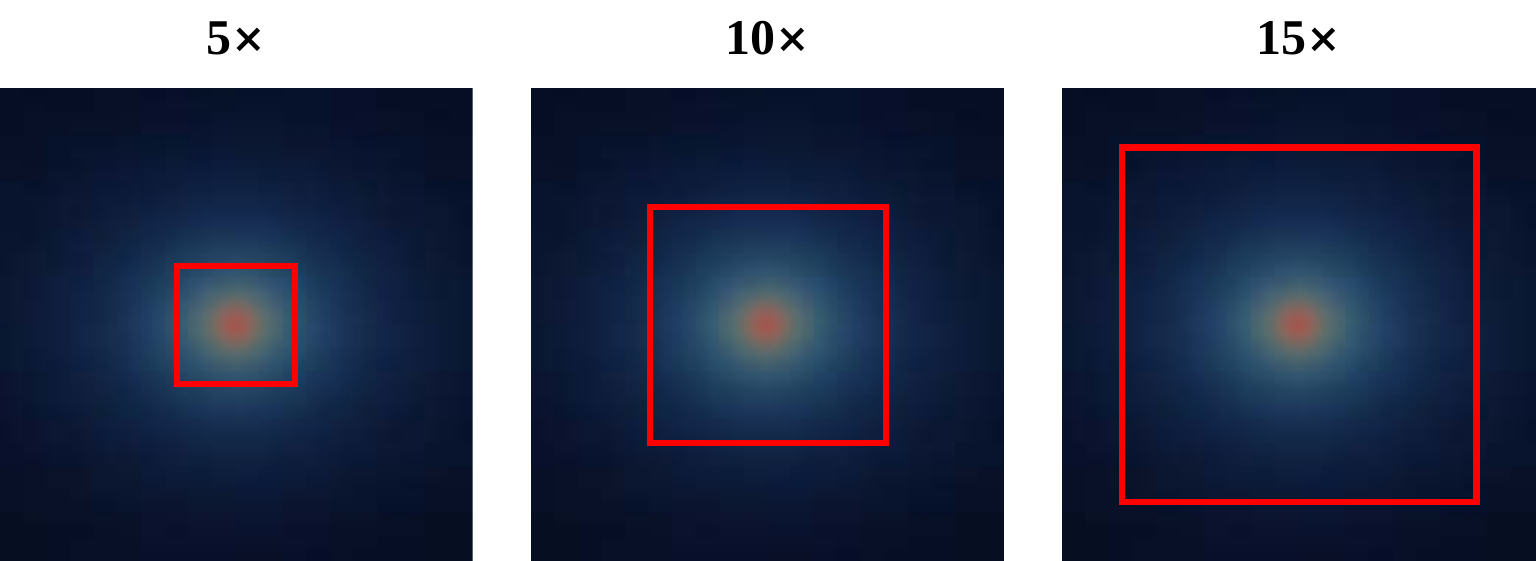}
    \caption{
    Pseudo GT BBoxes of different sizes on the average particle image.
    }
    \label{fig:heat_sc}
\end{figure}

To enhance accuracy, we propose incorporating contextual information of deposited energy to ViC.
This is motivated by the observation that the incident position is typically close to the clusters of activated cells, aligning with the diffusion characteristics of particles.
Consequently, we reformulate the incident position regression as an object detection problem, where the center of the predicted bounding box represents the incident position.
This allows us to leverage advanced object detection models and corresponding loss function designs to tackle the challenging task of accurately localizing incident particles.

The visual detection pipeline typically relies on ground-truth bounding box (GT BBox) annotations for training.
However, precise GT BBox annotations are not available for the particle incident images.
To address this limitation, we propose a pseudo bounding box (BBox) generation strategy which converts the pixel-level ground-truth incident positions to BBox representations.
Specifically, we treat the incident position as the center of the pseudo GT BBox and define its spatial extent as a multiple of the cell size in the image domain, Fig.~\ref{fig:pseudo_bbox}(a).
Through 2-D Gaussian fitting of the average particle image, it is calculated that the standard deviation $\sigma$ of the distribution is 14.53 pixels. Since the size of a single cell is approximately 8 pixels, 10$\times$ cell size is close to the $\pm3\sigma$ interval, which covers over 99\% of the energy within the average particle image, Fig.~\ref{fig:heat_sc}.
A smaller pseudo BBox imposes a stricter constraint on minimizing the prediction error but reduces the amount of contextual information about the deposited energy, while a large one captures more energy information without substantial additional gains, yet at the expense of localization precision.

\subsubsection{Incident Momentum Regression}

\begin{figure}[!t]
    \centering
    \includegraphics[width=1.0\linewidth]{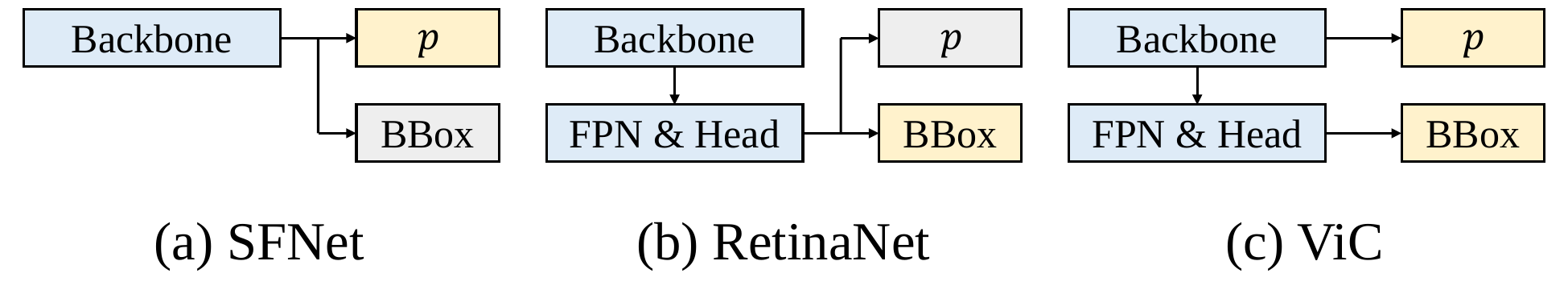}
    \caption{
    Comparison of regression structures in different object detectors.
    (a) SFNet regresses directly from backbone features, achieving good momentum but poor position accuracy.
    (b) RetinaNet regresses from detection head features, yielding good position but poor momentum.
    (c) Our ViC combines both strengths, using backbone features for momentum and detection head features for position, achieving high accuracy in both tasks.
    }
    \label{fig:sfnet}
\end{figure}

For position prediction, the locality of particle incidence and the distribution of activated cells provide sufficient information, while global information could introduce noisy information.
For momentum regression, in contrast, each energy deposition reflects momentum of the incident $\bar{n}$, and all depositions potentially carry meaningful information.
Therefore, the network needs a wider receptive field to capture global features for accurate estimation.
Accordingly, we maintain global attention for momentum regression while retaining local attention for position prediction, Fig.~\ref{fig:sfnet}.
The effectiveness of this design is validated through experiments.

To ensure that the regressed value of $p_{pred}$ is positive, we follow the design of BBox prediction decoder in visual object detectors~\cite{retinanet}.
Specifically, the calculation of $p_{pred}$ is performed as
\begin{equation}
    p_{pred} = p_{base} \cdot e ^ {p_{out}},
\end{equation}
where $p_{out}$ denotes the output of the momentum regression head.
$p_{base}$ is set to 1.0 based on the prior knowledge that $\bar{n}$ momentum in the dataset is concentrated near 1 GeV/$c$.

\section{Experiment}

\subsection{Experimental Setting}

\subsubsection{Dataset}

We collect a total of 986,343 electron-positron collision events from high-energy collision experiments conducted on BESIII~\cite{Asner:2008nq} and construct the \textbf{$\bar{n}$-1m} dataset.
The EMC responses of $\bar{n}$ are converted into 2-D images ($i.e.$, particle images) using the previously described method.
Each image is annotated with the incident position and momentum, forming the foundation for all $\bar{n}$ detection experiments.
To evaluate the model, we randomly select 100,000 samples as the testing set, while the remaining samples are used for training.

\subsubsection{Experimental Setup}

We adopt the standard protocol established in the MS COCO dataset~\cite{mscoco, mmdetection}, including the data augmentation that only use random horizontal flipping with a probability of 0.5, as well as the image pre-processing for color normalization using the statistical results from ImageNet~\cite{imagenet}, which has been widely used for training visual models~\cite{fasterrcnn, maskrcnn, retinanet}.
To adapt the visual object detector to particle images, we perform the following pre-processing steps: (i) resizing input image resolution to (960, 480), and (ii) retaining only the bounding box with the highest confidence for inference.
All backbone networks and FPNs follow their original hyperparameter settings.
All models are trained for 12 epochs using 8$\times$RTX 4090 GPUs with the total batch size of 128.
Except for ResNet-based models that use SGD optimizer with an initial learning rate of $1\times10^{-2}$, other models adopt the AdamW optimizer~\cite{adamw} with an initial learning rate of $1\times10^{-4}$ and a weight decay of $0.05$, and the learning rate would decay to 1/10 of the previous level after the 8th and 11th epochs.
When training ViC, these settings would consume 11,202 MB of memory per GPU and 18.5 hours of time-cost for about 886K $\bar{n}$ samples, while 4,450 MB of memory per GPU and 5 minutes of time-cost for 100K $\bar{n}$ samples when testing.
Unless otherwise stated in the ablation study, all backbones are pre-trained on ImageNet~\cite{imagenet}, and all experiments utilize pseudo GT BBoxes with dimensions set to 10$\times$ the size of an EMC cell.

\begin{figure*}[!t]
    \centering
    \includegraphics[width=0.9\linewidth]{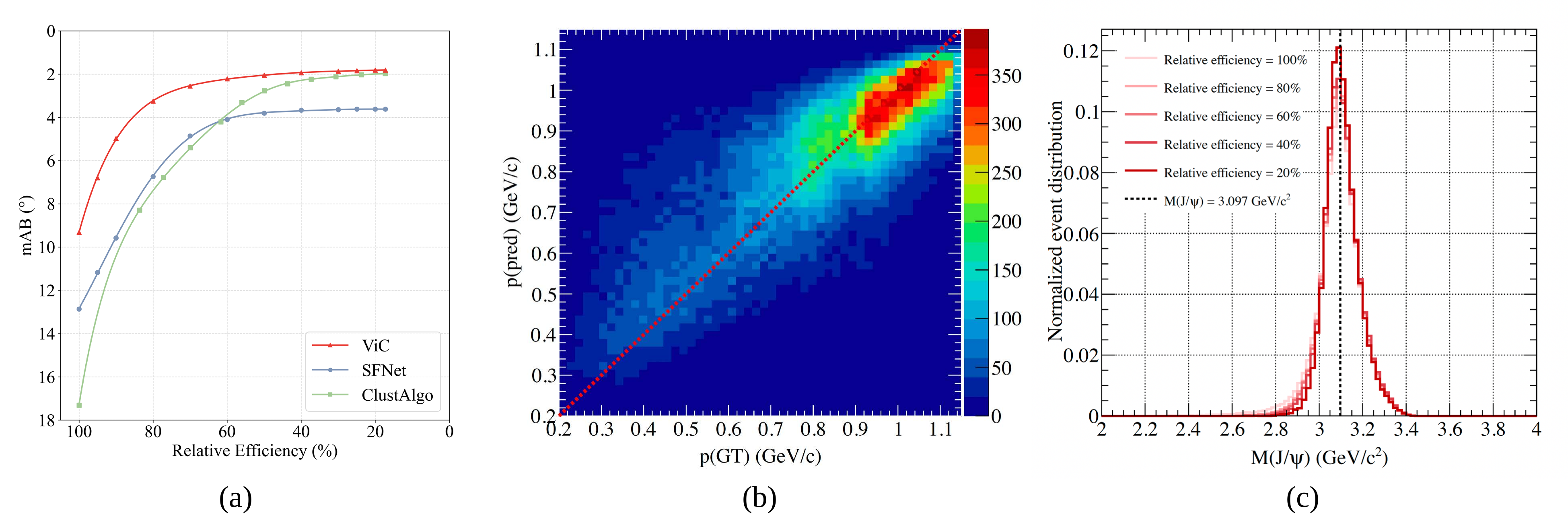}
    \caption{
    Performance of $\bar{n}$ parameter estimation with ViC.
    (a) mAB at relative efficiency levels;
    (b) Scatter plot of momentum predictions corresponding to their ground-truths;
    (c) Reconstructed $J/\psi$ peak in invariant mass spectrum based on $\bar{n}$ detection results.
    }
    \label{fig:results_Nm}
\end{figure*}

\begin{table*}[!t]
\centering
\caption{Performance comparison of networks.}
\begin{tabular}[t]{@{}c|c|cccc|ccc|cc@{}}
\toprule
\multirow{2}*{Method} & \multirow{2}*{ClustAlgo} & \multicolumn{4}{c|}{HEP ML Methods} & \multicolumn{3}{c|}{VODs (with vHeatK-T)} & \multirow{2}*{SFNet} & \multirow{2}*{ViC (ours)} \\
&& Set Trans. & EFN & ParticleNet & ParT & RetinaNet & Mask RCNN & DINO && \\
\midrule
mAB (°)  $\downarrow$ & 17.31 & 24.91 & 21.24 & 15.12 & 25.88 &  9.37 & 11.95 & 10.24 & 12.87 & \textbf{9.32} \\
mRE (\%) $\downarrow$ & -     & 32.30 & 35.62 & 30.75 & 34.60 & 22.28 & 31.99 & 22.20 & 22.04 & \textbf{21.48} \\
\bottomrule
\end{tabular}
\label{tab:performance_main}
\end{table*}

\subsubsection{Evaluation Metric}

As the final goal is to obtain the position and momentum of $\bar{n}$, we propose to quantify the error as the included angle between the ground-truth coordinates $(\phi_1, \theta_1)$ and the predicted coordinates $(\phi_2, \theta_2)$ in the spherical coordinate system.
This error, referred to as the angular bias (AB, noted as ${\gamma}$ in Fig.~\ref{fig:pseudo_bbox}(b)), is computed as
\begin{equation}
    \begin{aligned}
    \vec{v_i} &= (\cos{\phi_i}\sin{\theta_i}, \sin{\phi_i}\sin{\theta_i}, -\cos{\theta_i}), i\in\{1, 2\}, \\
    \mathrm{AB} &= \arccos{(\vec{v_1} \cdot \vec{v_2})}. \\
    \end{aligned}
    \label{eq:angular_bias}
\end{equation}
For position evaluation, we report the mean angular bias $m\mathrm{AB} = \frac{1}{N} \sum_{i=1}^{N} \mathrm{AB}_i$ on the entire testing set with $N$ samples, as well as on subsets of test samples with predicted confidence (derived from the classification score of foreground given by the object detector) above a specified threshold.
The ratio of the subset size to the size of the entire testing set is referred to as the \textit{relative efficiency}~\footnote{As physical observations rely on statistics of collision events, it is a common treatment~\cite{Asner:2008nq} to discard lower-quality events for higher precision.}.
For momentum evaluation, we compare $p_{pred}$ with $p_{gt}$ to calculate the relative error (RE) in percentage
\begin{equation}
    \mathrm{RE}=\frac{\left|p_{pred}-p_{gt}\right|}{p_{gt}}\times100\%.
\end{equation}
and report the mean relative error $m\mathrm{RE} = \frac{1}{N} \sum_{i=1}^{N} \mathrm{RE}_i$ upon the entire testing set with $N$ samples.

\subsection{Quantitative Results}

\subsubsection{Incident Position Prediction}

As shown in Table~\ref{tab:performance_main}, ViC achieves a smaller incident position prediction error, measured as the mean angular bias (mAB), compared with machine learning (ML) methods in the field of high-energy physics (HEP) including the ClustAlgo~\cite{He:2011zzd}, Set Transformer~\cite{lee2019set}, EFN~\cite{komiske2019energy}, ParticleNet~\cite{qu2020jet}, ParT~\cite{qu2022particle}, visual object detectors (VODs) including RetinaNet~\cite{retinanet}, Mask RCNN~\cite{maskrcnn}, and DINO~\cite{dino}, and our proposed point-wise regression solution (SFNet).
While the HEP-oriented ML methods exhibit a degree of universality when transferred to the $\bar{n}$-detection task, the VODs more effectively align natural image features with particle image representations.
ViC builds on these strengths and further adapts the architecture to HEP-specific constraints, yielding superior performance.

In Fig.~\ref{fig:results_Nm}(a), we also compare the mAB metric of ViC against ClustAlgo and SFNet under various relative efficiency settings.
At 100\% relative efficiency, where all predictions are included in the error calculation, ViC achieves a substantial improvement over ClustAlgo, reducing the prediction error by 46.16\% (from 17.31° to 9.32°).
Furthermore, ViC demonstrates a significant improvement over SFNet, underscoring the effectiveness of the proposed pseudo BBox generation strategy tailored for ViC.

At extremely low relative efficiency ($i.e.$, when only the highest-confidence predictions are retained), mAB approaches a lower bound of 1.8°.
Two main factors are considered to contribute to this lower bound:
(i) the spatial granularity of EMC cells, with the open angle of a single EMC cell ranging from 3° to 5.625°;
(ii) the inherent labeling uncertainty of 0.61° which is stated in the preliminary section.

\subsubsection{Incident Momentum Regression}

As shown in Table~\ref{tab:performance_main}, ViC also achieves a smaller incident momentum prediction error (mRE).
Specifically, the relative error of momentum regression on the entire dataset is 21.48\%.
Notably, ViC enables the BESIII EMC to outperform other calorimeters in particle physics experiments, even those specified for measuring the $\bar{n}$ momentum.
For comparison, the mRE of hadronic calorimeters exceeds 50\% in sub-GeV energy regions~\cite{Golutvin:2000ua}.
Furthermore, as shown in Fig.~\ref{fig:results_Nm}(b), the bright band in the scatter plot closely follows the diagonal line, which meet the requirement of statistical unbiasedness in physics, highlighting ViC's great potential.

\begin{figure*}[t]
    \centering
    \includegraphics[width=0.9\linewidth]{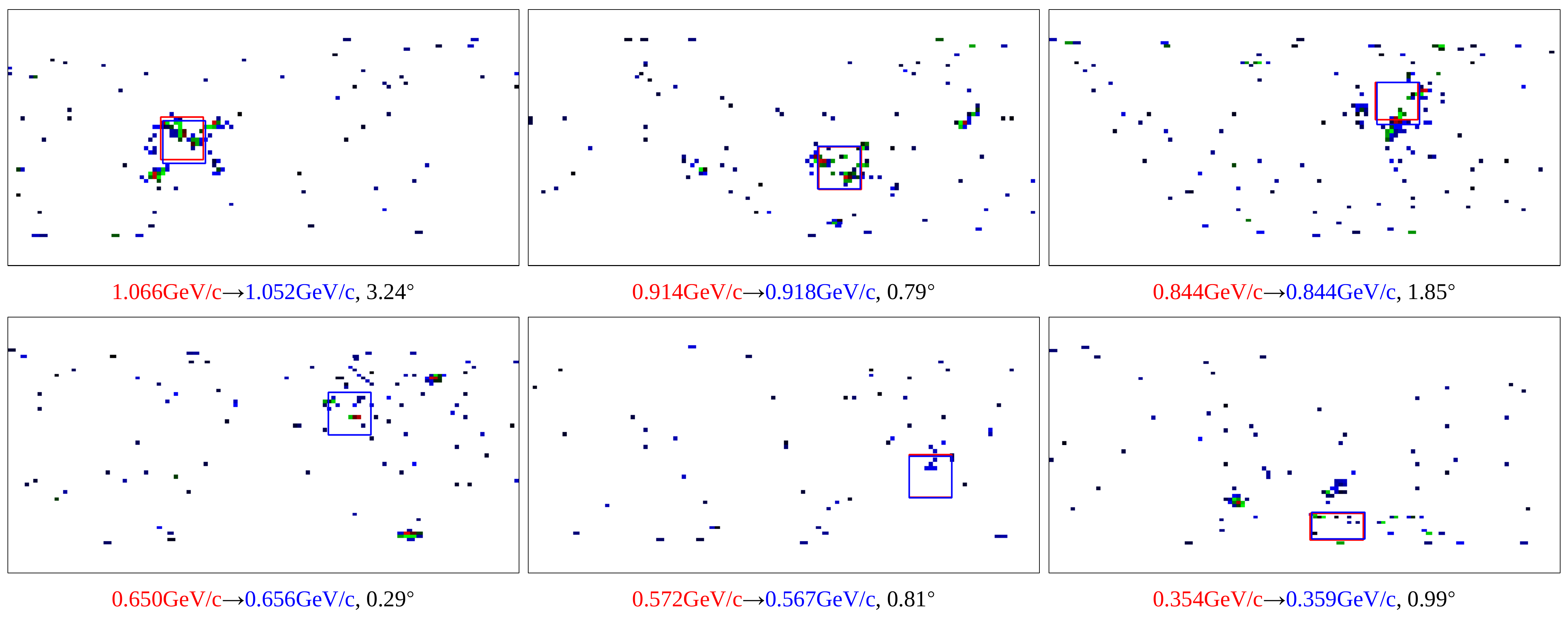}
    \caption{
    Visualization of $\bar{n}$ parameter estimation by ViC. Ground-truths are in red and predictions are in blue. The angular biases between ground-truths and predictions on each particle image are also marked.
    }
    \label{fig:show_Nm}
\end{figure*}

Fig.~\ref{fig:show_Nm} visualizes some results of $\bar{n}$ detection.
ViC consistently achieves accurate results in both position and momentum regression.
This validates that ViC has effectively generalized the pre-trained visual model to high-energy particle images with complex patterns.

\subsubsection{Capability in Physical Analysis}

The capabilities of $\bar{n}$ detection by ViC is to aid high-energy physics research, $e.g.$, hierarchical reconstruction of other particles decaying to $\bar{n}$.
Combining the $\bar{n}$ position and momentum predicted by ViC with those of other final-state particles, an important characteristic of the initial particle called \textit{invariant mass} can be calculated~\footnote{The discovery of the Higgs boson, for example, is represented as a peak in the invariant mass distribution of two photons.}.

We test the reconstruction task of $J/\psi$ particle decaying to a proton, a pion and the $\bar{n}$. As shown in Fig.~\ref{fig:results_Nm}(c), a sharp peak is formed in the invariant mass distribution, whose mean value correctly represents the $J/\psi$ nominal mass~\cite{ParticleDataGroup:2024cfk}.
The width of the peak can be reduced under lower relative efficiency settings, resulting in higher sharpness which is crucial for physical analysis.

\subsection{Ablation Study}

\subsubsection{Visual Representation Models}

\begin{table}[t]
\centering
\setlength{\tabcolsep}{0.15cm}
\caption{Performance comparison with different backbones.}
\begin{tabular}[t]{@{}c|ccccc@{}}
\toprule
\multirow{2}*{Backbone} & \multicolumn{5}{c}{Universal Representation Models (Scratch)} \\
                        & ResNet-50 & ConvNX-T & Swin-T & vHeat-T & vHeatK-T \\
\midrule
mAB (°)  $\downarrow$ & 14.07 & 10.08 & 10.73 &  9.69 &  9.51 \\
mRE (\%) $\downarrow$ & 37.80 & 22.97 & 23.47 & 22.41 & 21.88 \\
\midrule
\midrule
\multirow{2}*{Backbone} & \multicolumn{5}{c}{Universal Representation Models (Pre-trained)} \\
                        & ResNet-50 & ConvNX-T & Swin-T & vHeat-T & vHeatK-T \\
\midrule
mAB (°)  $\downarrow$ & 11.95 & 10.04 & 10.27 &  9.50 &  \textbf{9.32} \\
mRE (\%) $\downarrow$ & 34.71 & 22.47 & 22.40 & 21.63 & \textbf{21.48} \\
\bottomrule
\end{tabular}
\label{tab:performance_backbones}
\end{table}

\begin{figure*}[t]
    \centering
    \includegraphics[width=0.9\linewidth]{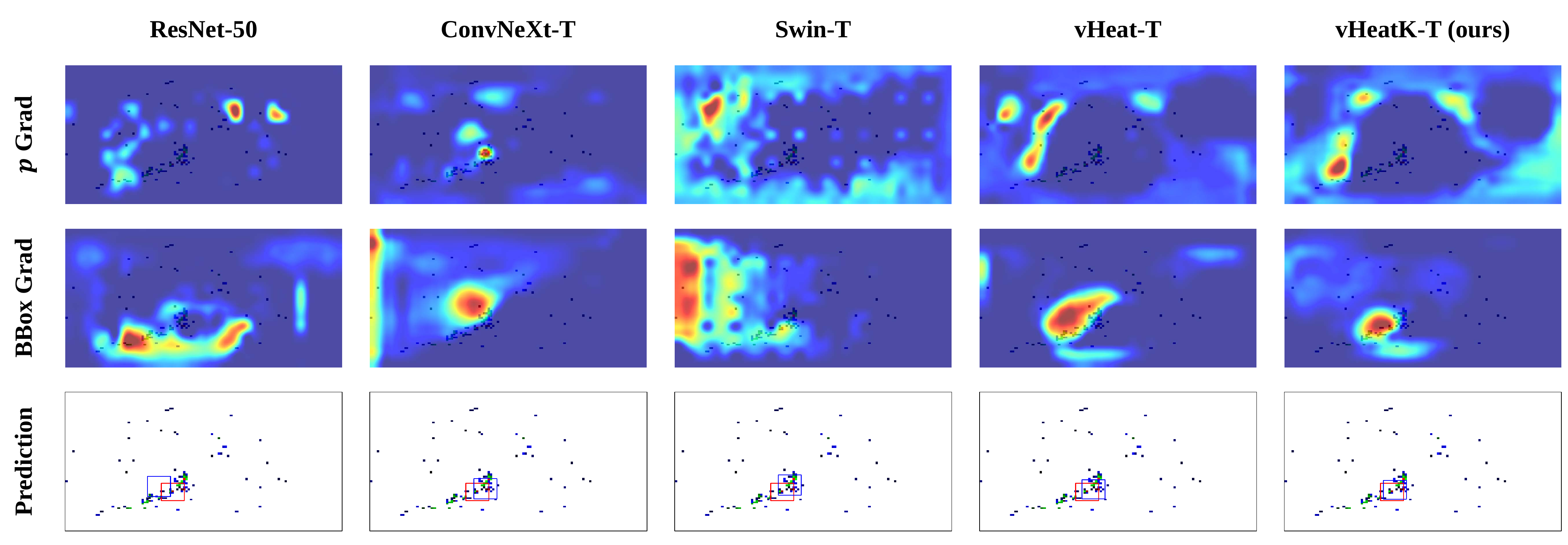}
    \caption{
    Visualization of gradient contributions using Grad-CAM~\cite{gradcam}. The proposed vHeatK combines global attention for momentum regression and local attention for position prediction, showing better prediction results than other models.
    }
    \label{fig:gradcam}
\end{figure*}

To demonstrate the effectiveness of the proposed vHeatK, we compare the performance of ViC with various universal representation models (ResNet~\cite{resnet}, ConvNeXt~\cite{convnext}, Swin Transformer~\cite{swin}, and the original vHeat~\cite{vheat2025}).
For a fair comparison, all models used in the experiments are adjusted to maintain a parameter range between 25M and 30M, and FLOPs between 4.1G and 4.6G, ensuring comparable model scales.
As shown in Table~\ref{tab:performance_backbones}, our vHeatK outperforms other benchmark backbone models in all scenarios.
Furthermore, as illustrated in Fig.~\ref{fig:gradcam}, vHeatK effectively combines local attention for position prediction and global attention for momentum regression, surpassing ConvNeXt, Swin Transformer, and its original version, vHeat.
The use of pre-trained models from the ImageNet~\cite{imagenet} vision dataset further enhances performance across most detectors, highlighting the potential for cross-domain alignment.

\begin{figure*}[t]
    \centering
    \includegraphics[width=0.9\linewidth]{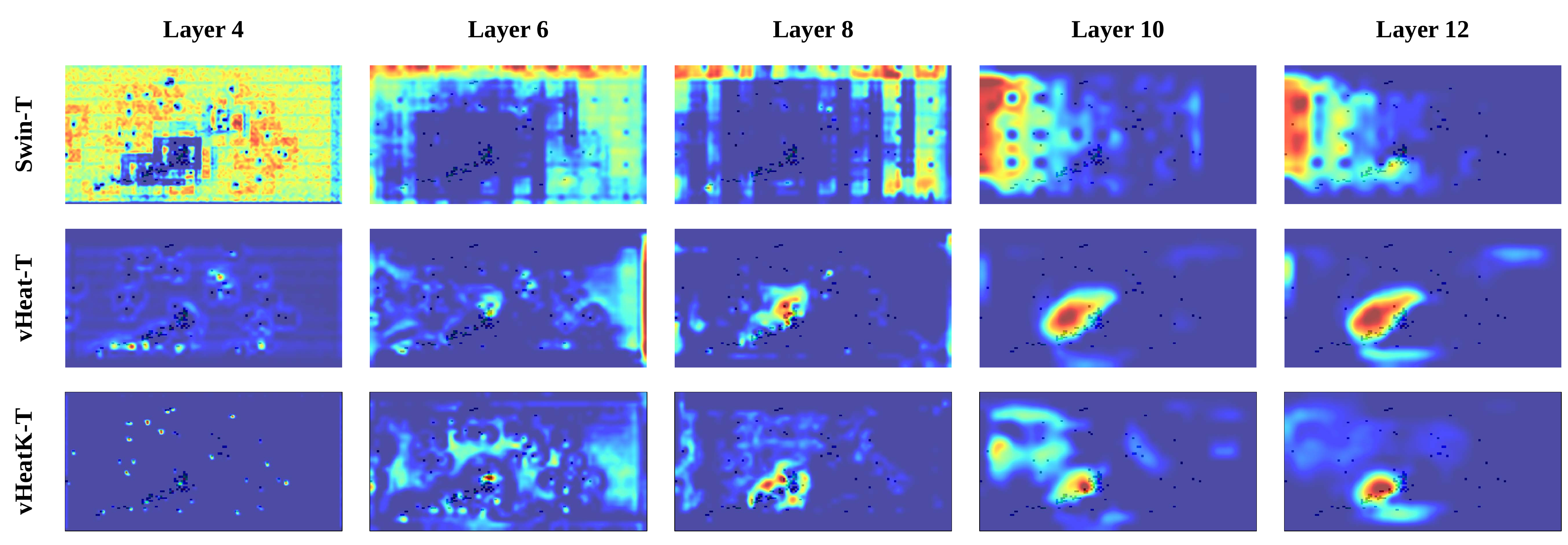}
    \caption{
    Visualization of attention maps from different layers using Grad-CAM~\cite{gradcam}.
    }
    \label{fig:gradcam_b}
\end{figure*}

We also visualize the attention maps derived from different layers.
As shown in Fig.~\ref{fig:gradcam_b}, unlike Swin Transformer based on self-attention operators whose attention is misguided by dominant background features most of the time, our heat-conduction-based network draws attention to foreground features already in the shallow layers.
The attention then diffuses outward just like heat-conduction, before gradually focusing on foreground regions in deeper layers. This process vividly illustrates how the heat-conduction operator facilitates feature extraction in a manner analogous to physical heat-conduction.
Furthermore, compared to the original vHeat model, our improved vHeatK architecture incorporates an additional feature fusion branch, which enhances the network's ability to adapt the heat-conduction behavior to different layers ($i.e.$, different stages of feature extraction). As a result, the attention becomes more focused and accurate across the hierarchy.

\subsubsection{Momentum Regression Head}

\begin{table}[t]
\centering
\caption{Evaluation of operators in the detector head. Conv. and Attn. denote 2-D convolution and self-attention, respectively.}
\begin{tabular}[t]{@{}c|ccc|c@{}}
\toprule
Operator & Conv. & Attn. & HCO & HCO-K (ours) \\
\midrule
mAB (°)  $\downarrow$ &  9.39 &  9.36 &  9.36 & \textbf{9.32} \\
mRE (\%) $\downarrow$ & 21.58 & 21.54 & 21.53 & \textbf{21.48} \\
\bottomrule
\end{tabular}
\label{tab:abla_operators}
\end{table}

\begin{table}[t]
\centering
\caption{Evaluation of layer numbers in the detector head.}
\begin{tabular}[t]{@{}c|ccccc@{}}
\toprule
Num of Layers & 0 & 1 & 2 & 3 & 4 \\
\midrule
mAB (°)  $\downarrow$ & 9.36  & 9.35  & \textbf{9.32}  & 9.40  & 9.42 \\
mRE (\%) $\downarrow$ & 21.59 & 21.56 & \textbf{21.48} & 21.52 & 21.65 \\
\bottomrule
\end{tabular}
\label{tab:abla_layers}
\end{table}

We compare the feature extraction operators in the detection head, including 2-D convolution~\cite{resnet}, self-attention~\cite{vit}, HCO~\cite{vheat2025}, and the proposed HCO-K.
As shown in Table~\ref{tab:abla_operators}, even the regression head does not use pre-training, HCO-K demonstrates a slight advantage over other operators.
Furthermore, we examine the impact of stacking numbers of HCO-K layers in the regression head.
Experimental results in Table~\ref{tab:abla_layers} indicate that a 2-layer configuration achieves the best performance.

\subsubsection{Size of Pseudo GT BBox}

\begin{table}[t]
\centering
\caption{Evaluation of pseudo GT BBox sizes.}
\begin{tabular}[t]{@{}c|ccc||ccc@{}}
\toprule
& \multicolumn{3}{c||}{Adaptive Size} & \multicolumn{3}{c}{Fixed Size} \\
& 5$\times$ & 10$\times$ & 15$\times$ & 40$^2$ & 80$^2$ & 120$^2$ \\
\midrule
mAB (°)  $\downarrow$ & 9.40  & \textbf{9.32}  & 9.52  & 9.42  & 9.40 & 9.49 \\
mRE (\%) $\downarrow$ & 21.58 & 21.48 & 21.42 & 21.56 & 21.49 & \textbf{21.38} \\
\bottomrule
\end{tabular}
\label{tab:abla_sc}
\end{table}

\begin{figure*}[t]
    \centering
    \includegraphics[width=0.9\linewidth]{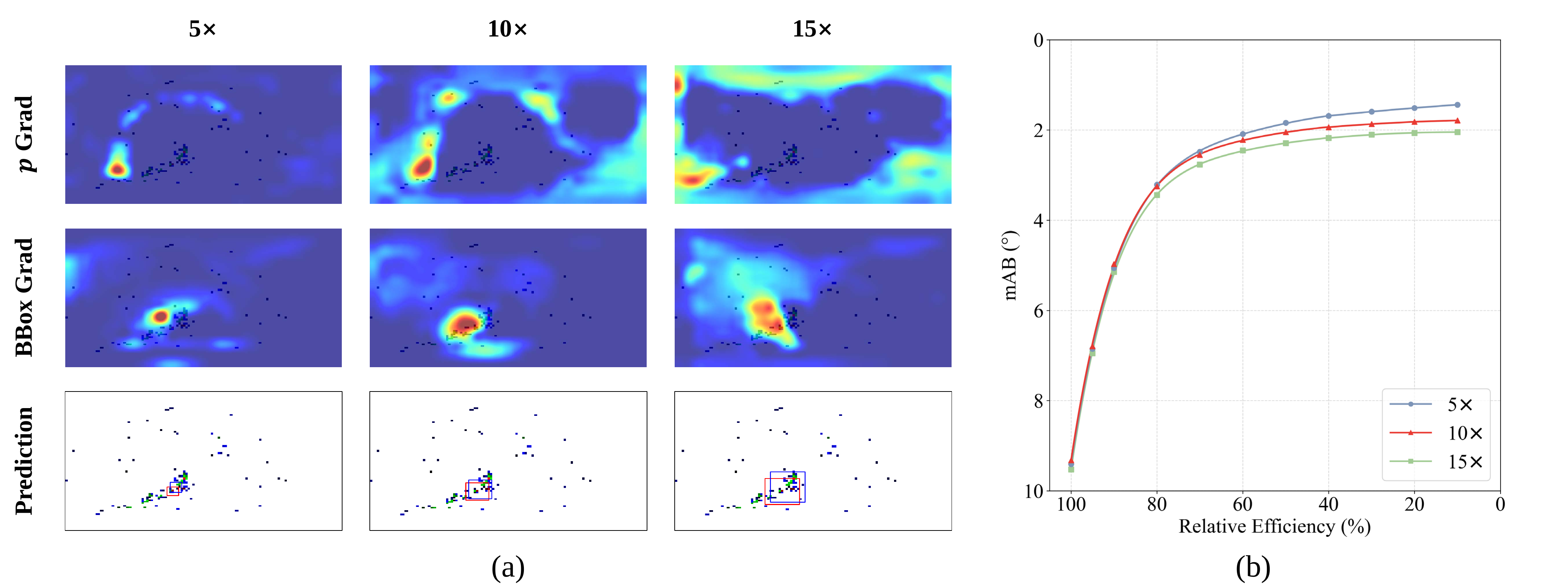}
    \caption{
    Evaluation of pseudo GT BBox sizes.
    (a) Gradient visualization using Grad-CAM~\cite{gradcam}. 
    While 5$\times$ size of the EMC cell shows a much smaller attention and 15$\times$ a much larger one, 10$\times$ shows a proper attention range.
    (b) Comparison of mAB at relative efficiency levels with different sizes of pseudo bounding boxes.
    }
    \label{fig:gradcam_sc}
\end{figure*}

This is critical for estimation accuracy, as it determines how much contextual information is included, Fig.~\ref{fig:gradcam_sc}(a).
We evaluate six pseudo GT BBox sizes: adaptive sizes, such as 5$\times$, 10$\times$, and 15$\times$ the size of the calorimeter cell at the incident position, and fixed sizes, including 40$\times$40, 80$\times$80, and 120$\times$120.
As shown in Table~\ref{tab:abla_sc}, the optimal performance under the position metric is achieved with a 10$\times$ size, while the momentum metric favors a 120$\times$120 size.
It is conclude that smaller pseudo boxes facilitate position prediction, whereas larger ones are more suitable for momentum regression.
Additionally, Fig.~\ref{fig:gradcam_sc}(b) shows that smaller sizes (e.g., 5$\times$) yield lower prediction errors in low relative efficiency scenarios, indicating that the optimal choice depends on specific task requirements.

\begin{figure*}[t]
    \centering
    \includegraphics[width=0.9\linewidth]{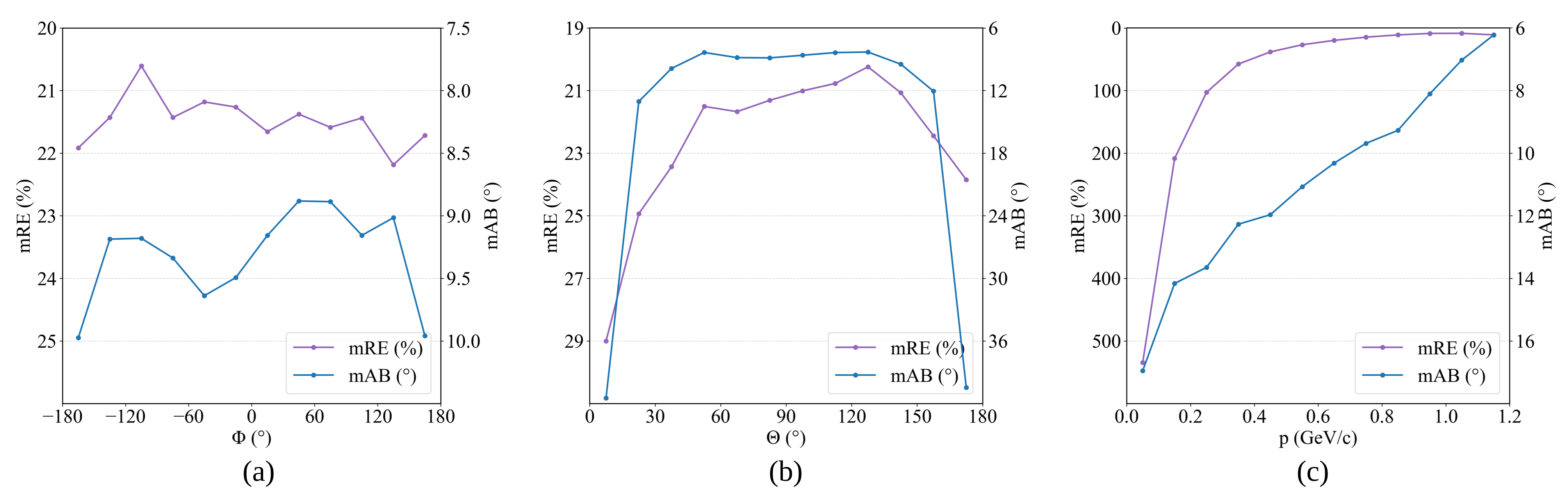}
    \caption{
    Evaluation of GT values.
    (a) the azimuthal angle $\phi$,
    (b) the polar angle $\theta$, and
    (c) the momentum $p$.
    }
    \label{fig:analysis_Nm}
\end{figure*}

\section{Conclusion}

We propose Vision Calorimeter (ViC), an end-to-end deep learning approach for anti-neutron ($\bar{n}$) detection using data from the electromagnetic calorimeter (EMC) cell arrays.
Leveraging particle images derived from EMC responses, ViC addresses the discrete and radial nature of particle images while employing a detection structure designed to capture both local and global contextual information embedded in the energy distribution.
Experimental results validate the effectiveness of ViC, outperforming the conventional methods and pioneering the capability of predicting incident momentum, for the first time.

We believe the significance of ViC lies in revealing the effectiveness and potential of adapting advanced machine learning technology to natural science domains represented by high-energy physics research.
Under the specific conditions of the BESIII experiment, ViC only needs to consider particles with momenta below 1.2 GeV/$c$, and benefits from the absence of pile-up in the experimental environment, which allows the method's adaptability to be clearly demonstrated.
Due to data access policies of other high-energy physics experiments, extending our approach across experiments will be an important direction for future work.

\section*{Acknowledgments}

We thank Zhaozhi Wang, Yuyang Huang and Zhi Cao for suggestions on heat-conduction models, Yajun Mao, Yangheng Zheng and Zhaoyang Yuan for discussion of high-energy physics. We thank BESIII Collaboration for their support on particle datasets. This study was supported by CAS Project for Young Scientists in Basic Research under Grant 14070, National Natural Science Foundation of China (NSFC) under Grant 62225208 and 62450046, and Zhongguancun Academy Project No.20240312.

\appendix
\subsection{Deduction of HCO Calculation Expression}

2-D heat-conduction follows the following equation, as
\begin{equation}
    \frac{\partial u}{\partial t} = k \left( \frac{\partial^2 u}{\partial x^2} + \frac{\partial^2 u}{\partial y^2} \right),
    \label{eq:heat}
\end{equation}
where $k$ is the heat conductivity. Setting the initial condition $f(x, y) = u(x, y, t) \big|_{t=0}$ and applying the Fourier Transform ($\mathcal{F}$), Eq.~\ref{eq:heat} has a solution, as
\begin{equation}
    \tilde{u}(\omega_x, \omega_y, t) = \tilde{f}(\omega_x, \omega_y)e^{-k(\omega_x^2 + \omega_y^2)t},
    \label{eq:heat_solve1}
\end{equation}
where
\begin{equation}
    \begin{aligned}
    \tilde{u}(\omega_x, \omega_y, t) & = \mathcal{F}(u(x, y, t)), \\
    \tilde{f}(\omega_x, \omega_y)    & = \mathcal{F}(f(x, y)). \\
    \end{aligned}
\end{equation}
Performing Inverse Fourier Transformer ($\mathcal{F}^{-1}$) on Eq.~\ref{eq:heat_solve1}, we have
\begin{equation}
      u(x, y, t)
    = \mathcal{F}^{-1}\left( \tilde{f}(\omega_x, \omega_y)e^{-k(\omega_x^2 + \omega_y^2)t} \right).
    \label{eq:heat_solve2}
\end{equation}
When applied to 2-D images, the (Inverse) Fourier Transform should be written in the form of (Inverse) Discrete Cosine Transform (DCT/IDCT). Therefore, Eq.~\ref{eq:heat_solve2} is converted to
\begin{equation}
    U^t = \mathbf{\mathrm{IDCT}}_\mathbf{\mathrm{2D}}\left( \mathbf{\mathrm{DCT}}_\mathbf{\mathrm{2D}}(U^0)e^{-k(\omega_x^2 + \omega_y^2)t} \right),
    \label{eq:hco}
\end{equation}
where $U^t$ denotes the feature map after heat-conduction time $t$. This expression is exactly the core calculation formula of HCO~\cite{vheat2025}.

\subsection{Error Analysis}

In Fig.~\ref{fig:analysis_Nm}, we summarize the dependencies of mAB on the azimuthal angle $\phi$, polar angle $\theta$, and momentum $p$.
At the left/right image boundaries of particle images, errors are typically 0.5°$\sim$1° higher than the central area (panel~(a)), which shows that the pseudo BBox approach used in ViC inherits the edge-error effect of BBox in visual object detectors.
Besides, errors increase by about 30° near the top/bottom of images (panel~(b)), owing both to visual object detectors' limitation and to the much lower cell density on the EMC end-caps compared with the barrel, which results in sparser deposited energy measurements.
In addition, higher-momentum $\bar{n}$ events yield more accurate position estimation (panel~(c)), which is consistent with physical prior: larger momentum implies greater EMC energy deposition, therefore, more reliable detection by ViC.

We further assess how mRE varies with $\phi$, $\theta$, and $p$ in Fig.~\ref{fig:analysis_Nm}.
Near the left/right image boundaries, momentum errors are slightly larger than that in the central region (panel~(a)), and they increase markedly at the top/bottom boundaries (panel~(b)), reflecting the detector-dominated deficiency.
Additionally, lower-momentum $\bar{n}$ events have less accurate regression (panel~(c)), which indicates that ample research room exists for ViC in the low-$p$ region in the future work.

\subsection{Model Generalization}

\begin{table}[t]
\centering
\caption{Comparison of training/testing sets from different decay processes.}
\begin{tabular}{cc|cccc}
\toprule
\multirow{3}*{\diagbox[width=72pt, height=27pt]{Testing Set}{Training Set}} 
& $\bar{n}$-1m     & \yesyes & \yesyes & \yesyes & \yesyes \\
& $\bar{n}$-lmdlmd &         & \yesyes &         & \yesyes \\
& $\bar{n}$-sgmsgm &         &         & \yesyes & \yesyes \\
\midrule
\multirow{2}*{$\bar{n}$-1m}
& mAB (°)  $\downarrow$ &  \textbf{9.32} &  9.43 &  9.44 &  9.41 \\
& mRE (\%) $\downarrow$ & \textbf{21.48} & 21.60 & 21.61 & 21.57 \\
\midrule
\multirow{2}*{$\bar{n}$-lmdlmd}
& mAB (°)  $\downarrow$ &  8.36 &  \textbf{5.38} & 11.75 &  \underline{5.76} \\
& mRE (\%) $\downarrow$ & 15.54 &  \textbf{3.49} & 14.01 &  \underline{3.67} \\
\midrule
\multirow{2}*{$\bar{n}$-sgmsgm}
& mAB (°)  $\downarrow$ & 25.39 & 28.86 & \textbf{11.40} & \underline{11.80} \\
& mRE (\%) $\downarrow$ & 20.40 & 22.00 &  \textbf{7.36} &  \underline{8.00} \\
\bottomrule
\end{tabular}
\label{tab:abla_sgmlmd}
\end{table}

\begin{figure}[t]
    \centering
    \includegraphics[width=1.0\linewidth]{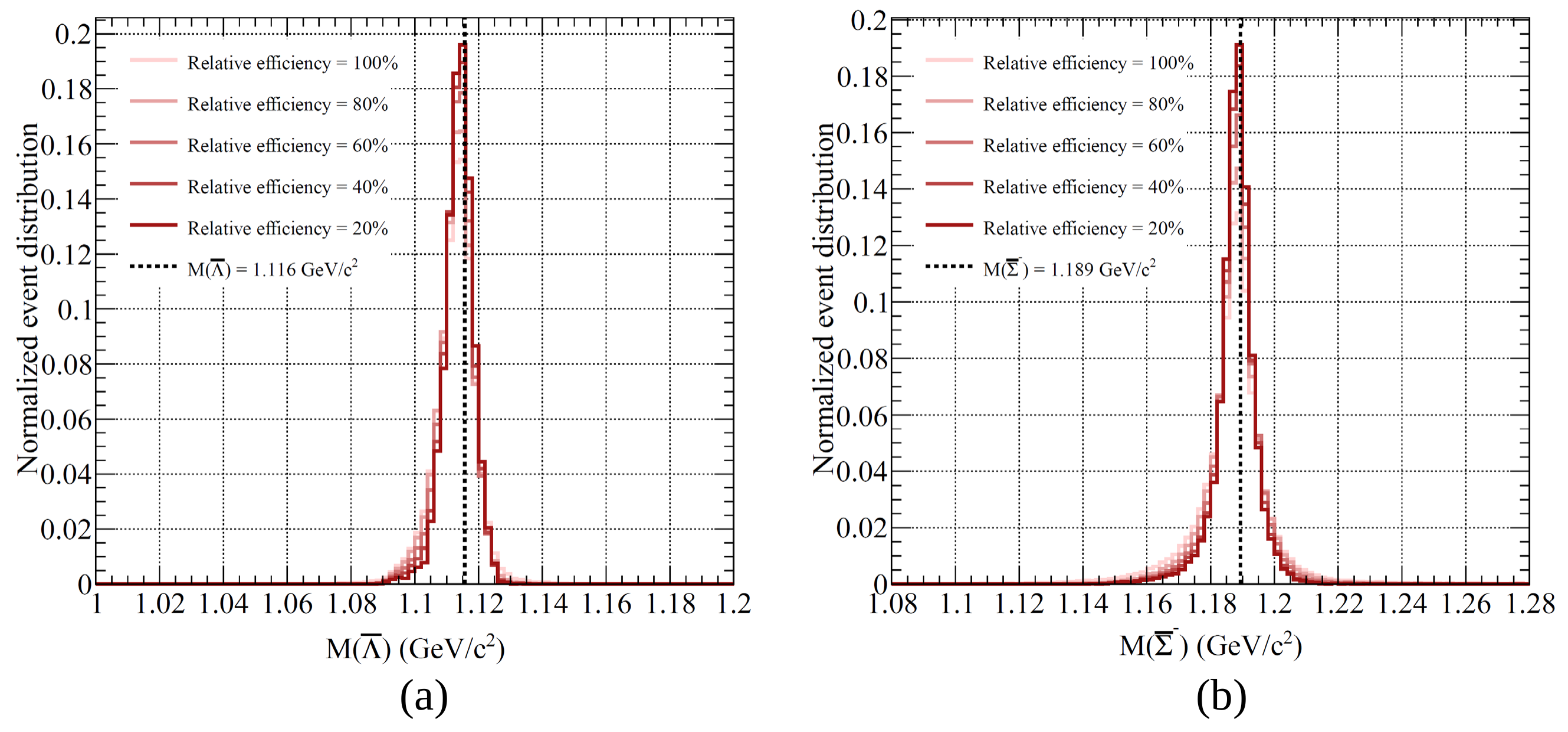}
    \caption{Reconstructed (a) $\bar{\Lambda}$ / (b) $\bar{\Sigma}$ peak in invariant mass spectrum.}
    \label{fig:sgmlmd_peak}
\end{figure}

We evaluate the generalization capacity of ViC on other decay processes containing $\bar{n}$.
The $\bar{n}$ datasets generated from $J/\psi{\rightarrow}\Lambda(p\pi^{-})\bar{\Lambda}(\bar{n}{\pi}^{0})$ and $J/\psi{\rightarrow}{\Sigma}^{+}(p\pi^{0}){\bar{\Sigma}}^{-}(\bar{n}{\pi}^{-})$ process are introduced, which contain 206,879 and 205,239 samples respectively, labeled as \textbf{$\bar{n}$-lmdlmd} and \textbf{$\bar{n}$-sgmsgm}.
Similar to the settings of $\bar{n}$-1m dataset, we randomly select 100,000 samples as the testing set, while the remaining samples are randomly mixed with $\bar{n}$-1m dataset for training.
As shown in Table~\ref{tab:abla_sgmlmd}, mixing half of the samples from each of these two decay processes into $\bar{n}$-1m dataset can achieve good performance on all three datasets.
Additionally, accurate reconstructions in invariant mass spectrum are also achieved for both two decay processes, Fig.~\ref{fig:sgmlmd_peak}, which show ViC's potential for high-energy physics researches.

\begin{figure}[!t]
    \centering
    \includegraphics[width=1.0\linewidth]{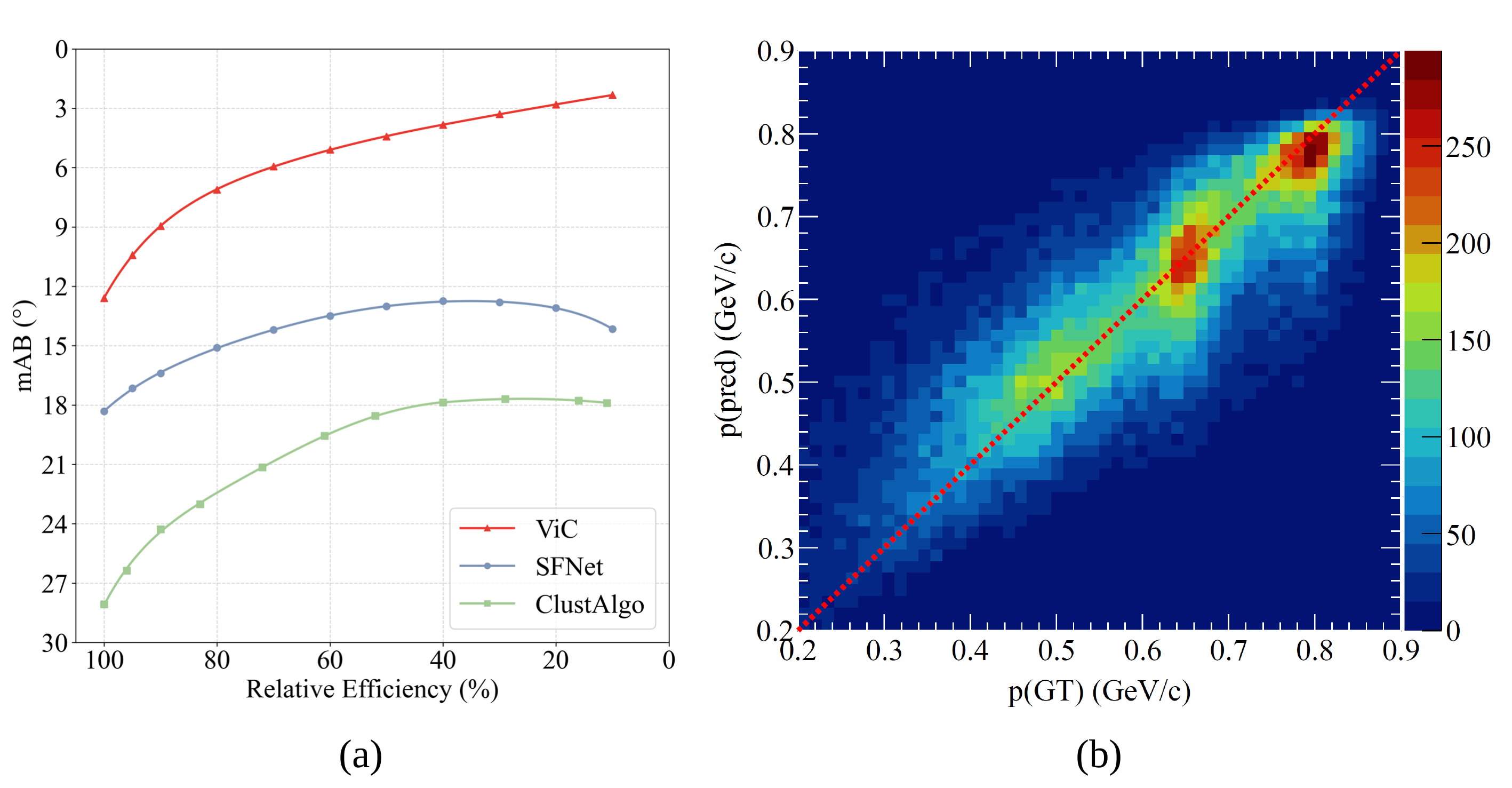}
    \caption{
    Performance of $\bar{\Lambda}$ detection with ViC.
    (a) mAB at relative efficiency levels;
    (b) Scatter plot of momentum predictions corresponding to their ground-truths.
    }
    \label{fig:results_Lmdm}
\end{figure}

To further show the generalization capacity, we extend ViC for the detection of other particle types, $e.g.$, anti-lambda ($\bar{\Lambda}$) particles~\footnote{The $\bar{\Lambda}$ particle is short-lived and eventually decay to an $\bar{n}$ and two photons before reaching the EMC. This results in image properties that are similar to, but distinct from, those of the $\bar{n}$ particle.}.
Specifically, we construct a training set consisting of 705,973 samples of $\bar{\Lambda}$ particles, along with a testing set of 80,000 samples.
Since we draw histogram of deposited energy, Fig.~\ref{fig:eng2rgb}(b), and find high similarity comparing to Fig.~\ref{fig:eng2rgb}(a, c, d), we believe the energy-to-color mapping expression Eq.~\ref{eq:eng2rgb} is also suitable for the $\bar{\Lambda}$ dataset.
As shown in Table~\ref{tab:abla_Lmdm} and Fig.~\ref{fig:results_Lmdm}, ViC achieves promising estimation performance with 12.59° mAB and 19.12\% mRE, outperforms the compared methods with significant margins.
This underscores ViC's generalization capability to a wider variety of particle types.

\begin{table}[!t]
\centering
\caption{Comparison of models on $\bar{\Lambda}$ dataset.}
\begin{tabular}[t]{@{}c|c|cc@{}}
\toprule
Detector & ClustAlgo & SFNet & ViC (ours) \\
\midrule
mAB (°)  $\downarrow$ & 28.07 & 18.32 & \textbf{12.59} \\
mRE (\%) $\downarrow$ & -     & 21.77 & \textbf{19.12} \\
\bottomrule
\end{tabular}
\label{tab:abla_Lmdm}
\end{table}

\bibliographystyle{IEEEtran}
\bibliography{bbbb}

\begin{IEEEbiography}[{\includegraphics[width=1in,height=1.25in,clip,keepaspectratio]{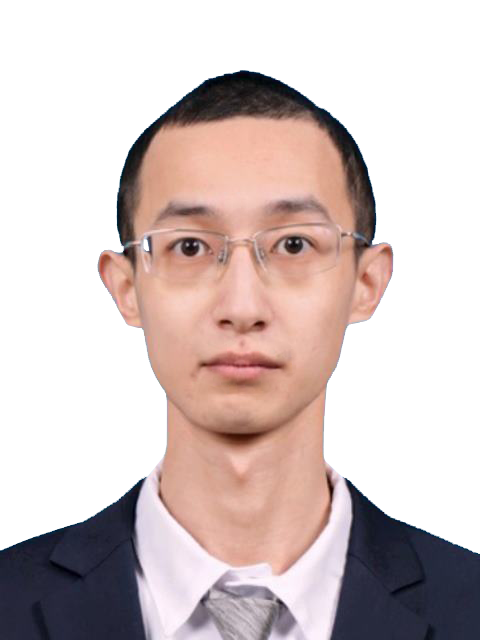}}]{Hongtian Yu} is a Ph.D. candidate of University of Chinese Academy of Sciences and Zhongguancun Academy, Beijing, China. He received the B.E. degree from University of Chinese Academy of Sciences in 2021. His research interests include visual object detection and AI for high-energy physics.
\end{IEEEbiography}

\begin{IEEEbiography}[{\includegraphics[width=1in,height=1.25in,clip,keepaspectratio]{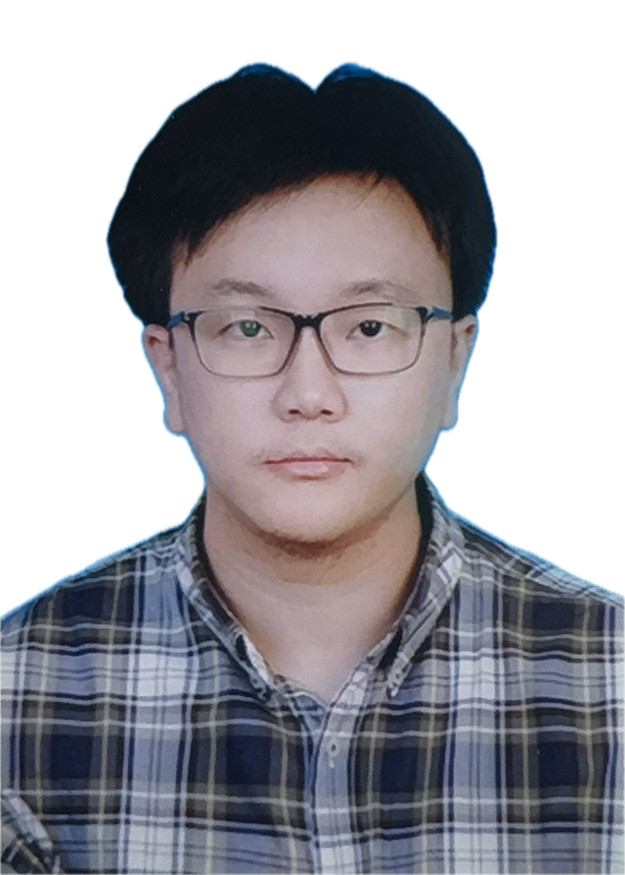}}]{Yangu Li} received the B.S. degree in Physics and the Ph.D. degree in Experimental Particle Physics from Peking University, China in 2020 and 2025. He is a postdoctoral fellow with the School of Physics, University of Chinese Academy of Sciences, China. His research interests include experimental measurements at particle colliders and the application of deep learning techniques in particle physics.
\end{IEEEbiography}

\begin{IEEEbiography}[{\includegraphics[width=1in,height=1.25in,clip,keepaspectratio]{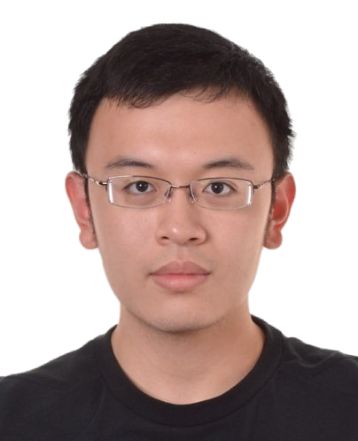}}] {Yunfan Liu} received the B.E. degree in electronic engineering from Tsinghua University, China in 2015, the M.S. degree in electronic engineering systems from University of Michigan, USA in 2017, and the Ph.D. degree from University of Chinese Academy of Sciences, China in 2023. He is a postdoctoral fellow with the University of Chinese Academy of Sciences, China. His research interests include computer vision and generative models.
\end{IEEEbiography}

\begin{IEEEbiography}[{\includegraphics[width=1in,height=1.25in,clip,keepaspectratio]{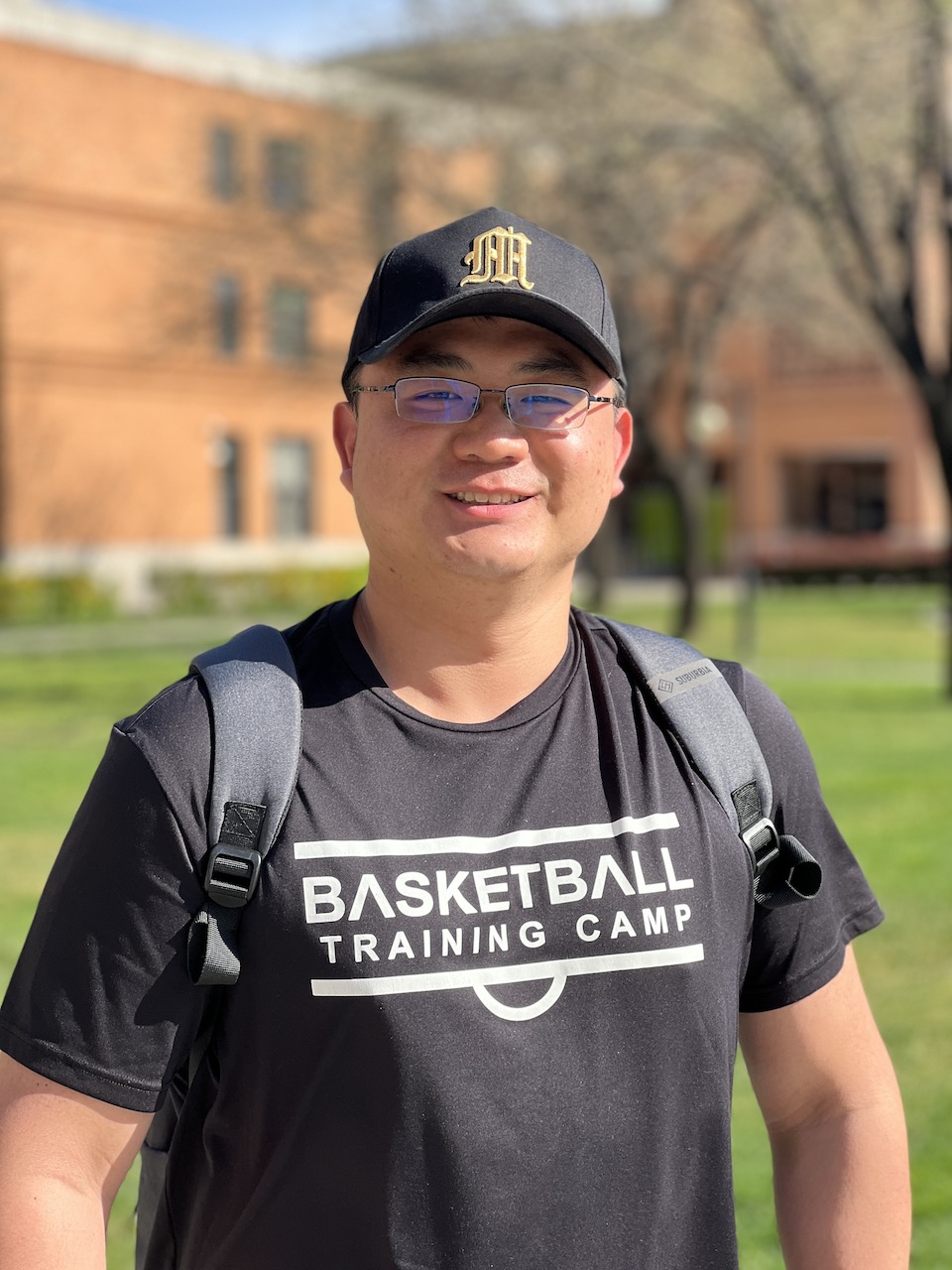}}]{Yunxuan Song} received the B.S. degree in Mathematics and Computing Sciences from Beijing Information Science and Technology University, China in 2017, and the Ph.D. degree in Experimental Particle Physics from Peking University, China in 2022. He is a postdoctoral fellow with the Le Laboratoire de Physique des Hautes Energies, École Polytechnique Fédérale de Lausanne, Switzerland. His research focuses on particle physics studies and exploring the use of deep learning methods in them.
\end{IEEEbiography}

\begin{IEEEbiography}[{\includegraphics[width=1in,height=1.25in,clip,keepaspectratio]{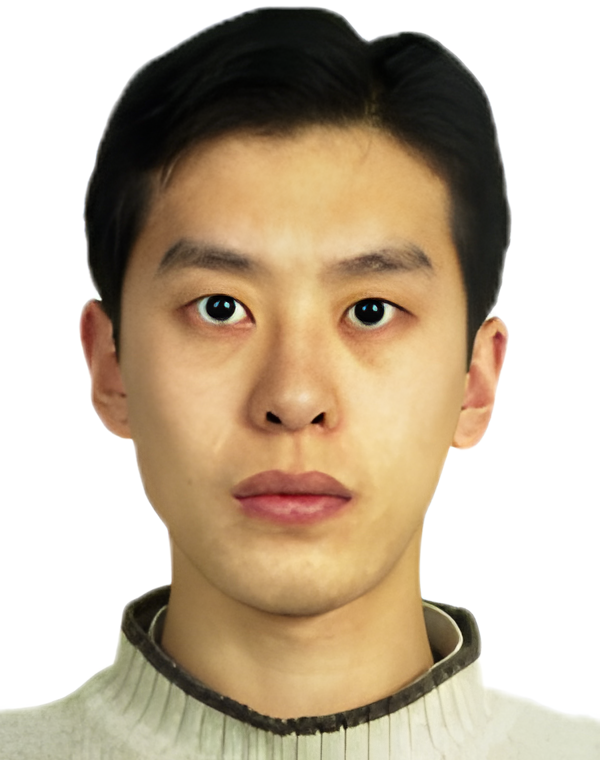}}]{Xiao-Rui Lyu} received the B.S. and M.S. degrees from Peking University, China in 2002 and 2005, respectively, and the Ph.D. degree from Tokyo Institute of Technology, Japan in 2008. He is a professor with the School of Physical Sciences, University of Chinese Academy of Sciences. His research focuses on experimental high-energy physics. He was the co-spokesperson and physics coordinator of the BESIII Collaboration, a large particle collider experiment which has over 600 members from 17 countries and regions, and has published more than 600 papers.
\end{IEEEbiography}

\begin{IEEEbiography}[{\includegraphics[width=1in,height=1.25in,clip,keepaspectratio]{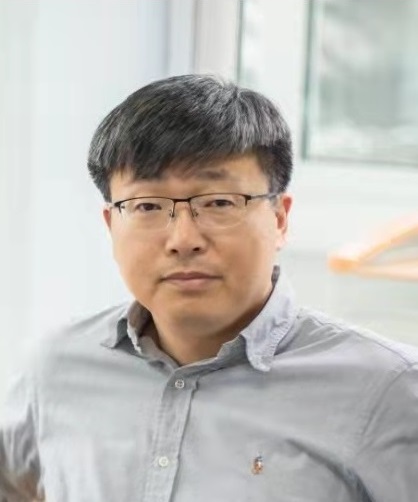}}]{Qixiang Ye}(M'10-SM'15) received the B.S. and M.S. degrees from Harbin Institute of Technology, China, in 1999 and 2001, respectively, and the Ph.D. degree from the Institute of Computing Technology, Chinese Academy of Sciences in 2006. He is a professor with the University of Chinese Academy of Sciences, and was a visiting assistant professor with the Institute of Advanced Computer Studies (UMIACS), University of Maryland, College Park until 2013. His research interests include image processing and machine learning. He has published more than 150 papers in refereed conferences and journals. He was on the editorial boards of IEEE Transactions on Circuit and Systems on Video Technology and IEEE Transactions on Intelligent Transportation Systems.
\end{IEEEbiography}

\end{document}